\documentclass[preprint,12pt,3p]{elsarticle}
\usepackage{graphicx}
\usepackage{amssymb}
\usepackage{amsmath}
\usepackage{makecell}
\usepackage{tikz}
\usepackage{dirtytalk}
\usepackage{float}
\usepackage{tabularx}
\usepackage{adjustbox}
\usetikzlibrary{shapes}
\usepackage{xcolor,colortbl}

\DeclareRobustCommand\sampleline[1]{%
\tikz\draw[#1] (0,0) (0,\the\dimexpr\fontdimen22\textfont2\relax)
-- (1.6em,\the\dimexpr\fontdimen22\textfont2\relax);}

\begin{document}
\begin{frontmatter}

\title{A general velocity correction scheme for two-way coupled point-particle simulations}

\author[label1]{Pedram Pakseresht}
\address[label1]{School of Mechanical, Industrial and Manufacturing Engineering, Oregon State University, Corvallis, OR 97331, USA}
\ead{pakserep@oregonstate.edu}
\author[label2]{Mahdi Esmaily}
\address[label2]{Sibley School of Mechanical and Aerospace Engineering, Cornell University, Ithaca, NY 14853, USA}
\ead{me399@cornell.edu}

%\author[label1]{Sourabh V. Apte}
\author[label1]{Sourabh V. Apte\corref{cor1}}
\ead{Sourabh.Apte@oregonstate.edu}

\cortext[cor1]{Corresponding author. 204 Rogers Hall, Corvallis, OR 97331, USA. Tel: +1 541 737 7335, Fax: +1 541 737 2600.}

%%%%%%%%%%%%%%%%%%%%
\begin{abstract}
The accuracy of Euler-Lagrange point-particle models employed in particle-laden fluid flow simulations depends on accurate estimation of the particle force through closure models. Typical force closure models require computation of the slip velocity at the particle location, which in turn requires accurate estimation of the {\it undisturbed} fluid velocity. However, when the fluid and particle phases are two-way coupled, wherein the particle and fluid phases exchange momentum through equal and opposite reaction forces, the fluid velocity field is disturbed by the presence of the particle. Since the undisturbed fluid velocity is not readily available, a common practice is to use the {\it disturbed} velocity, without any correction, to compute the particle force. This can result in errors as much as 100\% in predicting the particle dynamics. In this work, a general velocity correction scheme is developed that facilitates accurate estimation of the undisturbed fluid velocity in particle-laden fluid flows with and without no-slip walls. The model is generic and can handle particles of different size and density, arbitrary interpolation and distribution functions, anisotropic grids with large aspect ratios, and wall-bounded flows. The present correction scheme is motivated by the recent work of~Esmaily \& Horwitz (JCP, 2018) on unbounded particle-laden flows. Modifications necessary for wall-bounded flows are developed such that the undisturbed fluid velocity at any wall distance is accurately recovered, asymptotically approaching the unbounded scheme for particles far away from walls. A detailed series of verification tests were conducted on settling velocity of a particle in parallel and perpendicular motions to a no-slip wall. A range of flow parameters and grid configurations; involving anisotropic grids with aspect ratios typically encountered in particle-laden turbulent channel flows, were considered in detail. When the wall effects are accounted for, the present correction scheme reduces the errors in predicting the near-wall particle motion by one order of magnitude smaller values compared to the unbounded correction schemes. 

\end{abstract}
%%%%%%%%%%%%%%%

%%%%%%%%%%%%%%%%
\begin{keyword}
Wall-bounded particle-laden flows, Euler-Lagrange, Point-Particle models, Stokeslet solution, Wall effects.
\end{keyword}
%%%%%%%%%%%%%%%%

\end{frontmatter}

%%%%%%%%%%%%%%%%%%%%%%
\section{Introduction}
%%%%%%%%%%%%%%%%%%%%%%
%intro ...
Particle-laden flows are widely encountered in biology, nature and industry. Stroke by embolic particles in brain arteries \citep{mukherjee2016}, motion of red blood cells and margination of platelets in vessels \citep{muller2016}, drug delivery, urban pollutant and settling in human respiratory system, spray combustion \citep{apte2003}, particle-based solar receivers \citep{pouransari2017}, surgical site infection caused by dispersion of squames in the operating rooms \citep{he2018}, sediment transport \citep{finn2016}; among others are examples of such flows. Understanding the underlying physics of such flows, making predictions without performing expensive experiments, and ultimately optimizing the current systems require accurate predictive modelling tools.

The point-particle (PP) approach \citep{maxey1983,maxey1997} has received much attention in simulating these flows due to its simplicity, affordability and partial accuracy. This approach was initially introduced for modeling dilute particle-laden flows with relatively small size particles that have negligible effects on the background flow. For such a ``one-way" coupled flow \citep{elghobashi1991}, imposing the no-slip boundary condition on the surface of particles is not needed as the perturbation generated at the particle scale is insignificant. The fluid phase is solved using an Eulerian framework while particles are treated as Lagrangian points in the flow and tracked following the Newton's second law of motion based on the available closures for the fluid forces acting on the particles. Such one-way coupled simulations are mostly used for particle tracking and clustering. Nevertheless, owing to its affordability, this Euler-Lagrange (EL) approach has also been applied to particulate flows with dense loading or those with relatively large size particles wherein the effect of particles on the background flow is inevitable \citep{squires1990,elghobashi1993}. For such two-way coupled flows, the effect of particles on the carrier phase is modelled by applying the particle reaction force to the background flow through a momentum source term. Using such a simplified point force in modelling the inter-phase interactions, however, could result in some inaccuracies in capturing the experimental observations \citep{segura2005,eaton2009,pakseresht2014,pakseresht2015,pakseresht2016,pakseresht2017} or analytical solutions \citep{pan1996} of particle-laden flows.  

One source of inaccuracy is that, the fluid phase equations in this approach are solved for the entire flow field including the volumes occupied by the particles, and the mass displacement of the particles is not accounted for. Several works have shown the considerable effects of this displacement and have argued that this effect should be included in addition to the point-particle force \citep{ferrante2004,apte2008,cihonski2013,pakseresht2019}, in order to improve the predictions compared to the experimental observations. The other one, that is the focus of this work, is that the accuracy of PP in predicting the particle force can decay when the two phases are two-way coupled, owing to the disturbance created by the particle force on the background flow. Such a disturbance produces an error in the force calculations since the closure models often rely on the slip velocity computed based on the {\it undisturbed} fluid flow, which is not readily available in the two-way coupled simulations.  

Few schemes have been recently developed as a substitute for the standard PP approach in order to improve the modeling of particle-laden flows. \citet{pan1996} were the first to develop a velocity-disturbance-model that couples two phases through the velocity field rather than the momentum exchange force. Their model is based upon the Stokes solution for the motion of a particle in a quiescent flow, for which the flow field generated around the particle is analytically known. Accordingly, to couple the two phases and capture the particle's effect on the flow, one could directly enforce this solution to the background flow. Unlike the standard PP approach, this velocity-disturbance-model eliminates any dependency to the undisturbed fluid velocity and results in more accurate inter-phase coupling. Despite its accuracy, it is limited to flows with particles in the Stokesian regime ($Re_p{<}O(0.1)$). \cite{maxey2001} introduced an alternative scheme that approximately satisfies the no-slip boundary condition at the particle surface, that is suitable for particle-laden flows with relatively large particle sizes. In this force-coupling model, the presence of particles on the flow is approximated by a multipole expansion of a regularized steady Stokes solution. Despite its promising results for unbounded flows, for wall-bounded regimes it requires higher order terms, more than monopole and dipole, in order to accurately capture the wall lubrication effect \citep{lomholt2002} which in turn adds more complexity to their formulation. In addition, similar to \cite{pan1996} scheme, the assumption of Stokesian regime for flow around the particles limits the application of their method to flows where $Re_p{<}O(0.1)$.  

Recently, efforts have been made in order to improve the accuracy of the standard PP approach by retrieving the undisturbed fluid velocity from the available disturbed field. \cite{gualtieri2015} regularized the PP approach for the unbounded flows by deriving analytical equations to remove the self-induced velocity disturbance created by the particles. Their approach requires considerable computational resources to resolve the stencil over which the particle force is distributed using a Gaussian filter function. \cite{horwitz2016,horwitz2018} originated a method to obtain the undisturbed velocity based on the enhanced curvature in the disturbed velocity field for particle Reynolds numbers of $Re_p{<}10.0$. A C-field library data was built using reverse engineering technique that should be added to the current EL-PP approaches for recovering the undisturbed velocity. Although their model showed excellent agreement in the predictions of particle settling velocity and decaying isotropic turbulence \citep{mehrabadi2018}, it is limited to (i) the isotropic computational grids, (ii) particle-laden flows with particles with the maximum size of the grid ($\Lambda{=}d_p/\Delta$) of O(1), where $\Delta$ is the grid size and $d_p$ particle diameter, and (iii) the unbounded flows. \cite{ireland2017} derived an analytical expression for recovering the undisturbed velocity in unbounded flows based on the steady state Stokes solution that was derived as the solution of a feedback force distributed to the background flow using a Gaussian smoothening. Although their model accounts for the mass displacement of the particles, it is limited to unbounded flows with small $Re_p$. 

In a generic approach, \cite{esmaily2018} originated a correction scheme in which each computational cell is treated as a solid object that is immersed in the fluid. Each computational cell that is subjected to the two-way coupling force is dragged at a velocity that is identical to the disturbance created by the particle. In their physics-based model, the disturbance of each computational cell created by the particle is obtained by solving the Lagrangian motion of the cell concurrently with the equation of motion of the particle. Although their model was devised to handle (i) relatively large size particles ($\Lambda{>}1$), (ii) isotropic and anisotropic grids, (iii) flows with finite $Re_p$, and (iv) arbitrary interpolation and distribution functions, it is limited to unbounded flows. \cite{balachandar2019} developed a model based on analytical and empirical equations that correct the PP approach for modelling particle-laden flows with a wide range of particle Reynolds number, $Re_p{<}200$. Following their scheme, analogous model was developed by \cite{liu2019self} for retrieving the undisturbed temperature in heated particle-laden flows. Although their velocity and temperature models account for the mass displacement of the particles (similar to \cite{ireland2017}) and are built for a wide range of particle Reynolds number and Peclet number, they are derived for unbounded flows only, and based on a specific filter function; namely Gaussian, that limits their applicability. 

Nearly all available correction schemes have been originated and developed for the unbounded particle-laden flows. Due to the the wide range of wall-bounded applications, developing more general correction schemes that are applicable for flows near solid boundaries is necessary. \cite{pakseresht2019_aps} and \cite{horwitz2019_aps} underscored the need for such general correction schemes while \cite{battista2019} extended their regularized PP scheme \citep{gualtieri2015} for a turbulent particle-laden pipe flow. Unique modeling issues arise in wall-bounded particulate flows that need to be addressed in any correction scheme. First, particles near a wall, specially in a turbulent flow, are relatively bigger than the grid size normal to the wall and consequently disturb the flow strongly and anisotrpically. It has been observed that the disturbance created by a particle is proportional to the ratio of its volume to that of the cell \citep{esmaily2018}, hence the disturbance of particles near the wall is expected to be strong. Second, the correction scheme should be able to handle the anisotropic grid resolution typically encountered near the walls in turbulent particle-laden flows. Third, unlike unbounded flows, the disturbance created by a particle near the wall is conceptually asymmetric and should decay faster to the wall, in order to satisfy the no-slip boundary condition. These criteria necessitate the need for a general correction scheme that can capture any type of disturbance in presence or absence of the no-slip walls. 

This paper aims to develop such a generic correction scheme that meets the criteria, mentioned above. Such a scheme enables accurate predictions of wall-bounded, particle-laden flows, and will potentially help provide insights into the underlying physics of such flows. In this regard, the correction scheme originated by \cite{esmaily2018} (hereinafter named as E\&H) is generalized and extended to account for the wall effects on the disturbance field in the presence of no-slip boundary conditions. Additional adjustments are made due to the collocated grid arrangement used in this study. The generalized framework can be easily extended to complex arbitrary shaped, unstructured grids, as well as walls with curvature and surface roughness. The newly developed scheme is general and could be implemented and applied to all types of flows with different grid resolutions, arbitrary interpolation functions and varying particle to grid size. The new approach will be tested on canonical cases for which analytical solutions are available and illustrates the need for such a general correction scheme. How much the disturbance created by particle in the presence of no-slip wall gets deviated from its unbounded counterpart and how this affects the particle's motion and the inter-phase coupling in the presence of no-slip wall, are the questions that we address in this paper. 

%structure for paper and its sections ... 
The paper is organized as follows. We describe our correction scheme in section \ref{sec:scheme}. Correction factors due to the presence of a no-slip wall are introduced and the model is expanded to a wide range of grid resolutions typically encountered in wall-bounded turbulent particle-laden flows. Section \ref{sec:results} validates the model on predicting the velocity of a single particle settling in an unbounded domain. Then, the new model will be tested for velocity of a single particle moving parallel to the wall at various wall-normal distances. In addition, the perpendicular motion of a particle toward the wall is examined to assess the model for disturbances created in the wall-normal direction. In order to quantify the accuracy of the model for a wide range of applications, different flow parameters and computational grids are studied. Isotropic and anisotropic grid resolutions are investigated to demonstrate the capability of the model for different configurations. In order to illustrate the importance and the need for the present approach, the results are compared with the unbounded version of the present model, wherein wall effects are ignored, as well as the uncorrected scheme. Section \ref{sec:conclusion} concludes the paper with final remarks and summary of the work. 
%%%%%%%%%%%%%%%%%%%%%%%%%%%%%%%%%%%%%
\section{A general correction scheme}
%%%%%%%%%%%%%%%%%%%%%%%%%%%%%%%%%%%%%
\label{sec:scheme}
In this section, we first introduce the main underlying issue in the two-way coupled point-particle (PP) approach, then present a general methodology to resolve the issue in the presence and absence of the no-slip walls. In the standard PP approach, particles are assumed spherical and subgrid (smaller than the grid resolution), and tracked in a Lagrangian framework using the second law of Newton as, 

\begin{equation}
    m_p\frac{du^{(i)}_p}{dt} = F^{(i)} + m_pg^{(i)},
    \label{eq:newton_2ndlaw}
\end{equation}

\noindent wherein the particle velocity in direction $i$, $u^{(i)}_p$, with mass of $m_p$ is obtained using the total force of $F^{(i)}$ acting over the particle as well as its weight, $m_pg^{(i)}$. Depending upon the regime under consideration, different forces such as steady stokes drag ($F^{(i)}_d$), shear-induced lift ($F^{(i)}_l$), Magnus effect ($F^{(i)}_m$), buoyancy ($F^{(i)}_b$), added mass ($F^{(i)}_a$), history ($F^{(i)}_h$) and other forces may be included in the calculation of $F^{(i)}$,  

\begin{equation}
    F^{(i)} = F^{(i)}_d + F^{(i)}_b + F^{(i)}_a + F^{(i)}_h + F^{(i)}_l + F^{(i)}_m + ...,
\end{equation}

\noindent to accurately capture the motion of the particle \citep{maxey1983}. Most of these forces are derived for a setting in which the upstream flow field in known and unaffected by the presence of particle. As an example, the steady state Stokes drag force over a sphere with diameter of $d_p$ and in a fluid with dynamic viscosity of $\mu$ is

\begin{equation}
    F^{(i)}_d = 3\pi\mu d_p\left(u^{(i)}_f - u^{(i)}_p\right),
    \label{eq:stokes_drag}
\end{equation}

\noindent which is analytically derived based on the relative velocity between the {\it undisturbed} (upstream) fluid velocity, $u^{(i)}_f$, and the particle velocity of $u^{(i)}_p$. When two-phases are one-way coupled, i.e., the presence of particles do not affect the background flow through the momentum exchange \citep{elghobashi1991}, this force is employed for tracking the particle to obtain its velocity and position as a function of time. In such a scenario, the particle force is not exerted to the flow and the fluid phase remains undisturbed. This process yields an accurate (and consistent with the closure model) computation of $u^{(i)}_f$ and thereby Eqs. \ref{eq:newton_2ndlaw} and \ref{eq:stokes_drag}. In contrast, when the two phases are two-way coupled, this force, with the same magnitude and opposite direction, is applied back to the background flow to capture the inter-phase momentum interactions. This inter-phase coupling disturbs the fluid velocity around the particle and the newly disturbed velocity, $u^{(i)}_d$, that is different from the undisturbed velocity, $u^{(i)}_f$, is used in the calculation of the drag force for the next time step. This force computed based on the disturbed fluid velocity is inaccurate and yields erroneous trajectory of the particle as well as the inter-phase momentum interactions. For simple canonical particle-laden flows that are not bounded, this inaccuracy depends on flow parameters such as (i) particle diameter to the grid size ratio ($\Lambda$), (ii) the choice of interpolation and distribution functions used in the PP approach, (iii) particle Reynolds number and (iv) particle Stokes numbers \citep{horwitz2016,esmaily2018}. Computing the undisturbed fluid velocity might be easy for some simple flows such as settling of a particle in a quiescent flow, as the unaffected field could be readily obtained from the upstream condition. However, for more complex flows with large number of particles, particularly in wall-bounded regimes, such a naive remedy becomes invalid due to the fact that the whole flow field is disturbed. This issue necessitates development of a unified framework to accurately recover the undisturbed fluid velocity in general unbounded or wall-bounded particle-laden flows. The basic concept behind development of such a framework is described below. 

Since the disturbed fluid velocity in a two-way coupled PP approach arises from a point-force, finding the disturbance created by this force can be used to correct the disturbed flow and obtained the undisturbed fluid velocity. In other words, after a point force is applied to fluid within a computational cell in a discretized domain, what is the cell fluid velocity (let us denote it by $u^{(i)}_c$) generated by this force, and what does it depend upon are the main questions under consideration. The $u^{(i)}_c$ is the velocity that is missing in the traditional two-way coupled PP approaches, and if found, could be added to the disturbed fluid velocity to obtain the undisturbed velocity as 

\begin{equation}
    u^{(i)}_f = u^{(i)}_d - u^{(i)}_c
    \label{eq:uf}
\end{equation}

Thus, any predictive scheme that can model $u^{(i)}_c$, would be able to accurately recover the undisturbed fluid velocity. The correction scheme presented here is based on a method to predict the velocity of fluid in the computational cell produced by a force applied at its cell center. 

\begin{figure}
    \centering
    \includegraphics[scale=0.4]{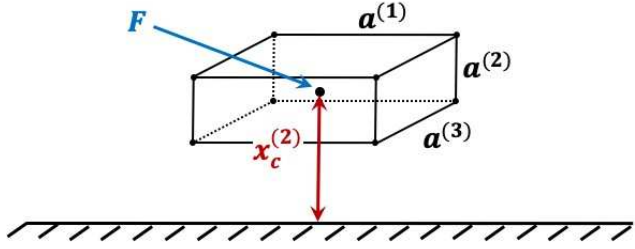}
    \caption{A computational cell with an arbitrary size of [$a^{(1)},a^{(2)},a^{(3)}$] and wall-normal distance of $x^{(2)}_c$ that is disturbed by force $\mathbf{F}$.}
    \label{fig:cell_near_wall}
\end{figure}

To obtain a generalized approach applicable to wide range of unbounded and wall-bounded particle-laden flows with different grid aspect ratios, consider a force of $F^{(i)}$ is applied to fluid at a computational cell in an anisotropic, Cartesian grid, that has an arbitrary size of $[a^{(1)},a^{(2)},a^{(3)}]$ and located near a no-slip wall, at a wall-normal distance of $x^{(2)}_c$ as shown in Fig. \ref{fig:cell_near_wall}. Also, suppose that the force is applied to the center of the computational cell, i.e., is generated by a particle that is at the center of the computational cell. Hereinafter, the superscripts $(1)$ and $(3)$ are employed for streamwise and spanwise directions, respectively, while, $(2)$ denotes the wall-normal direction. Note that here we use anisotropic Cartesian grids for simplicity, but this concept can be easily extended to arbitrary shaped unstructured grids with complex boundary walls as well, in the future. Conceptually, the time dependent velocity created by this force could be approximated as 

\begin{equation}
u^{(i)}_c(t) \approx f(a^{(1)},a^{(2)},a^{(3)},F^{(i)},t,x^{(2)}_c)
\end{equation}

By varying the grid aspect ratio, the distance to the wall, and the amount of point-force applied, a data-set for the disturbance velocity of the computational cell as a function of time can be generated. Although finding a generic function for this data set may require some advanced data-science techniques, this relationship can be significantly simplified by applying a small force that limits us to the creeping/Stokes flow regime. For a small force and in the steady state condition, the velocity of the computational cell is linearly proportional to the force, i.e., $u^{(i)}_c{\propto}F^{(i)}$, and one can write it as a function of the cell dimensions and its wall distance, i.e., $u^{(i)}_c{=}F^{(i)}g(a^{(1)},a^{(2)},a^{(3)},x^{(2)}_c)$. This hypothesis is examined to a computational cell with an arbitrary size and situated at a wall distance. A small force is applied to this cell and its velocity as a function of time is measured. Regardless of size and the location of the cell, it is observed that its velocity exponentially accelerates till reaches a terminal velocity, precisely similar to the settling velocity of a spherical particle under gravity and in the presence of a drag force. Motivated by this observation and following \cite{esmaily2018}, we model the computational cell as a solid object that is subjected to the particle force $F^{(i)}$, and dragged through the surrounding computational cells. At steady state, the particle force and the drag force exerted by the surrounding computational cells balance each other and the computational cell velocity becomes only a function of its size and wall distance. The general form of the model then can be written using a Maxey-Riely equation of motion for the computational cell velocity, including the unsteady effect as, 

\begin{equation}
    \frac{3}{2} m_c \frac{du^{(i)}_c}{dt} = -3 \pi \mu d_c K^{(i)}_t u^{(i)}_c - F^{(i)},
    \label{eq:uc}
\end{equation}

\noindent where $d_c$ is the volume-equivalent diameter of the computational cell ($d_c{=}\sqrt[3]{(6/\pi)a^{(1)}a^{(2)}a^{(3)}}$) with mass of $m_c{=}(\pi/6)\rho_fd^3_c$. The term on the left hand side expresses the unsteady effect of the force on the computational velocity wherein the prefactor $3/2$ captures the added mass effect. The first term on the right hand side of the equation, $3\pi \mu d_c K^{(i)}_t u^{(i)}_c$, is the Stokes drag force acting on the computational cell by its surrounding cells wherein the relative velocity is $-u^{(i)}_c$ as the ambient flow for the disturbance field is at rest. The adjustment to the Stokes drag is expressed by the factor $K^{(i)}_t$ as,  

\begin{equation}
    K^{(i)}_t = \frac{K^{(i)}_c C_r}{K^{(i)}_p C^{(i)}_t}.
    \label{eq:Kt}
\end{equation}

\noindent Here, $K^{(i)}_c$ accounts for non-sphericity of the computational cell and depends on its size and aspect ratio. The factor $K^{(i)}_p$ accounts for wall effects as well as the interpolation and distribution functions typically employed in PP approach. The factor $C_r$ accounts for the non-linear finite force effects whereas $C^{(i)}_t$ considers the limited exposure time of the particle force to the computational cell. These geometric and physics-based factors are defined and explained in details in the following subsection. 

%%%%%%%%%%%%%%%%%%%%%%%%%%%%%%%%%%%%%%%%%%%%%%%
\subsection{Geometric correction factor, $K_c$}
%%%%%%%%%%%%%%%%%%%%%%%%%%%%%%%%%%%%%%%%%%%%%%%
The geometric correction factor, $K_c$, is obtained based on the fact that a moving solid object in an unbounded flow with a small Reynolds number experiences a constant drag coefficient that is dependent on its shape and geometry \citep{leith1987}. Inspired by this, the geometric correction factor to the Stokes drag of the computational cell is conjectured to be a function of its size. In this part, an expression for $K_c$ is derived that is different than the one derived in the E\&H work, in order to cover a wider range of grid sizes and aspect ratios, typically encountered in highly turbulent particle-laden channel flows. 

The procedure is explained as follows. A sufficiently large computational domain is chosen with a uniform grid resolution of $128^3$. Boundary conditions for wall-normal direction are set to be no-slip and slip to enforce wall effects while periodic boundary condition is employed for the other directions of the domain. A small and stationary force, $F^{(i)}_{small}$, that generates a disturbance field with nearly zero Reynolds number, is applied to the center of a computational cell in $i$ direction. Note that the computational cell is located in the middle of a large domain wherein the no-slip boundary conditions have zero effect on the generated disturbance field. At steady state, the velocity of the computational cell is directly measured and $K^{(i)}_c$ is obtained by using Eq. \ref{eq:uc} as,

\begin{equation}
    K^{(i)}_{c,measured} = \left|\frac{F^{(i)}_{small}}{3 \pi \mu d_cu^{(i)}_c}\right|
    \label{eq:kc_measured}
\end{equation}

\noindent with other correction factors being one by definition as the force is small ($C_r{=}1$), applied only to one cell and sufficiently away from the no-slip wall ($K^{(i)}_p{=}1$), and has infinite exposure time ($C^{(i)}_t{=}1$). The procedure is repeated for a wide range of grid size of $0.05{\sim}a^{(2)}/a^{(1)}{\sim}1$ and $0.1{\sim}a^{(3)}/a^{(1)}{\sim}1$. The choice of grid size and aspect ratio studied here is inspired by the grid resolution of highly turbulent channel flows \citep{moser1999}. A best fit to the numerically measured data is obtained as,

\begin{equation}
\begin{split}
    K_c^{(i)} =&~ 0.1705\exp\left[(\Gamma_{max}^{(i)})^{-0.4005}(\Gamma_{min}^{(i)})^{0.06408}\right] (\Gamma_{max}^{(i)})^{0.7058}(\Gamma_{min}^{(i)})^{-0.452} \\ 
    + &\ln\left[(\Gamma_{max}^{(i)})^{-0.03746}(\Gamma_{min}^{(i)})^{0.2049}\right](\Gamma_{max}^{(i)})^{0.355}(\Gamma_{min}^{(i)})^{0.05338},
\end{split}
\label{eq:kc}
\end{equation}

\noindent where 
\begin{equation}
    \Gamma_{max}^{(i)} = \max \left\{\frac{a^{(j)}}{a^{(i)}},\frac{a^{(k)}}{a^{(i)}}\right\}, \quad \Gamma_{min}^{(i)} = \min \left\{\frac{a^{(j)}}{a^{(i)}},\frac{a^{(k)}}{a^{(i)}}\right\}; \quad j,k\neq i. 
    \label{eq:kc_cont}
\end{equation}

\begin{figure}
    \centering
    \includegraphics[scale=0.6]{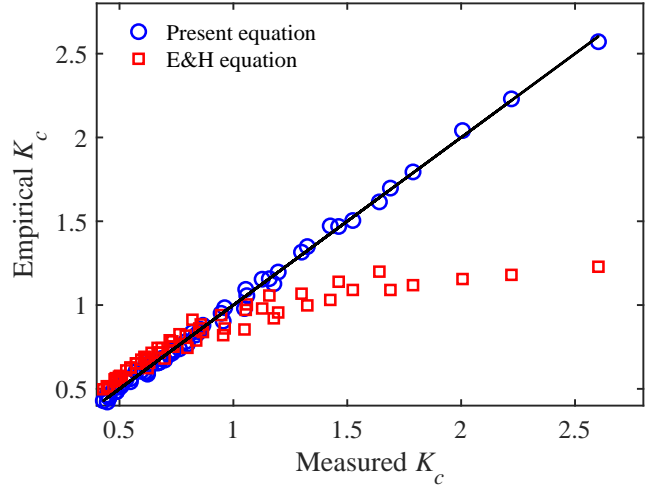}
    \caption{Prediction of Eq. \ref{eq:kc} versus numerical measurements of $K_c$ for a wide range of grid sizes typically encountered in wall-bounded turbulent channel flows.}
    \label{fig:kc_}
\end{figure}

Figure \ref{fig:kc_} shows excellent prediction of the above empirical equation against our numerical measurement for $K^{(i)}_c$. The prediction of the corresponding expression used in E\&H is also shown. For the studied range of grid sizes, our new correlation matches with E\&H for small $K^{(i)}_c$, but for larger $K^{(i)}_c$ values, which correspond to computational cells with higher aspect ratio, the new correlation matches much better than E\&H. In the next part, we show the derivation of wall effects as well as the interpolation effects. 

%%%%%%%%%%%%%%%%%%%%%%%%%%%%%%%%%%%%%%%%%%%%%%%%%%%%%%%
\subsection{The wall and interpolation effects}
%%%%%%%%%%%%%%%%%%%%%%%%%%%%%%%%%%%%%%%%%%%%%%%%%%%%%%%
The question that arises now is how does the geometric correction factor, $K^{(i)}_c$, change when the computational cell of interest gets closer to the wall? The answer for this question lies in a new wall adjustment factor on geometric correction factor. In order to answer this question we first look at the near wall motion of a spherical object wherein its drag coefficient increases as a function of wall distance. \cite{goldman1967quiscent} derived an analytical equation for the wall-modified drag coefficient of a sphere moving parallel to the wall, while \cite{brenner1961} using lubrication theory, obtained the corresponding parameter for its normal motion toward the wall. Based on these observations, it is expected that the wall adjustment on the geometric correction factor be dependent on the force direction and increases as wall-normal distance decreases. Having such a direction dependent adjustment is of importance as in wall-bounded particle-laden flows, particles interact with the sweep and burst events near the wall \citep{righetti2004}, thus experiencing different forces in the two directions and disturbing the background flow differently. 

Following the procedure described in the previous part for obtaining the $K^{(i)}_c$, its wall adjustment is achieved by applying the point-force at various wall distances. For each wall distance, Eq. \ref{eq:kc_measured} gives rise to a wall-modified geometric correction factor, $K^{(i)}_{c,w}$, that deviates from its unbounded counterpart, $K^{(i)}_{c}$. The ratio of these two yields a wall adjustment factor as

\begin{equation}
    \Psi^{(i)}_k = \frac{K^{(i)}_{c,w}}{K^{(i)}_c}.
    \label{eq:psi_c}
\end{equation}

This factor approaches unity for cells sufficiently away from the wall (i.e., $K^{(i)}_{c,w}=K^{(i)}_c$) and is greater than one for those near the wall. This procedure is repeated for the studied range of the grid resolutions, for each of which, $\Psi^{(i)}_k$ for various wall distances with both wall-normal as well as parallel forces were measured and tabulated. As explained in Appendix A, for isotropic grid resolution, it is observed that the wall adjustment to the Stokes drag coefficient of a spherical object obtained empirically by \cite{zeng2009} matches our measured data. This expression, however, deviates for highly skewed anisotrpic grids, inevitably encountered in the wall-bounded flows. This underscores the need for a more accurate expression that could handle a wide range of grid aspect ratios. The best fit to our measured data for forces in both parallel and normal directions was found to be,

\begin{equation}
    \Psi^{(i)}_{k} = 1 + \frac{A^{(i)}}{1+B^{(i)}h^{(i)}_k},
    \label{eq:psi}
\end{equation}

\noindent where $h^{(i)}_k$ is the normalized wall distance of the center of the computational cell of interest as 

\begin{equation}
    h^{(i)}_k = 
    \begin{cases}
    \frac{x^{(2)}_k}{a^{(i)}}, \quad \text{i=1,3}\\
    \frac{x^{(2)}_k}{a^{(1)}}, \quad \text{i=2}\\
    \end{cases}
    \label{eq:psi_cont1}
\end{equation}

\noindent with $x^{(2)}_k$ being the dimensional wall distance of the computational cell, and $A^{(i)}$ and $B^{(i)}$ are dependend on the grid size as, 

\begin{equation}
    A^{(i)} = 
    \begin{cases}
    \frac{\ln\left(26.31\frac{a^{(3)}}{a^{(1)}}\right)}{\left(0.05761+5.373\left(\frac{a^{(2)}}{a^{(1)}}\right)^{1.057}\right)}, \quad i=1 \\
    \frac{\ln\left(14.04\frac{a^{(3)}}{a^{(1)}}\right)}{\left(0.06608+5.14\left(\frac{a^{(2)}}{a^{(1)}}\right)^{1.592}\right)}, \quad i=2 \\
    \frac{\ln\left(26.31\frac{a^{(1)}}{a^{(3)}}\right)}{\left(0.05761+5.373\left(\frac{a^{(2)}}{a^{(3)}}\right)^{1.057}\right)}, \quad i=3 \\
    \end{cases}
     \label{eq:psi_cont2}
\end{equation}

\begin{equation}
    B^{(i)} = 
    \begin{cases}
    \frac{\exp\left(-0.02873\frac{a^{(3)}}{a^{(1)}}\right)}{\left(0.00008+0.5601\left(\frac{a^{(2)}}{a^{(1)}}\right)^{1.894}\right)}, \quad i=1 \\
    \frac{\exp\left(-1.252\frac{a^{(3)}}{a^{(1)}}\right)}{\left(0.01354+3.688\left(\frac{a^{(2)}}{a^{(1)}}\right)^{2.202}\right)}, \quad i=2 \\
    \frac{\exp\left(-0.02873\frac{a^{(1)}}{a^{(3)}}\right)}{\left(0.00008+0.5601\left(\frac{a^{(2)}}{a^{(3)}}\right)^{1.894}\right)}, \quad i=3 \\
    \end{cases}
    \label{eq:psi_cont3}
\end{equation}

As implied by Eq. \ref{eq:psi}, $\Psi^{(i)}_k$ becomes unity when the disturbance occurs sufficiently away from the wall as,  

\begin{equation}
    \lim_{h^{(i)}_k \to \infty} \Psi^{(i)}_k =1.
\end{equation}

%\noindent This choice of $\Psi^{(i)}_k$ is advantageous and makes the model general and applicable to all types of particle-laden flows.

It should be noted that our results show that for disturbances created by the wall-normal force applied to highly skewed grids, i.e., $a^{(2)}/a^{(3)}{<}0.5$, $\Psi^{(i)}_k$ for the first computational cell attached to the wall is better predicted by,

\begin{equation}
    \Psi^{(2)}_{first,cell} = \frac{\ln\left(25.3 \frac{a^{(3)}}{a^{(1)}} \right)}{-0.0007149+2.364\left( \frac{a^{(2)}}{a^{(1)}}\right)^{0.7796}}.
\label{eq:psi_normal_cl_wall}
\end{equation}

\begin{figure}
    \centering
    \includegraphics[scale=0.55]{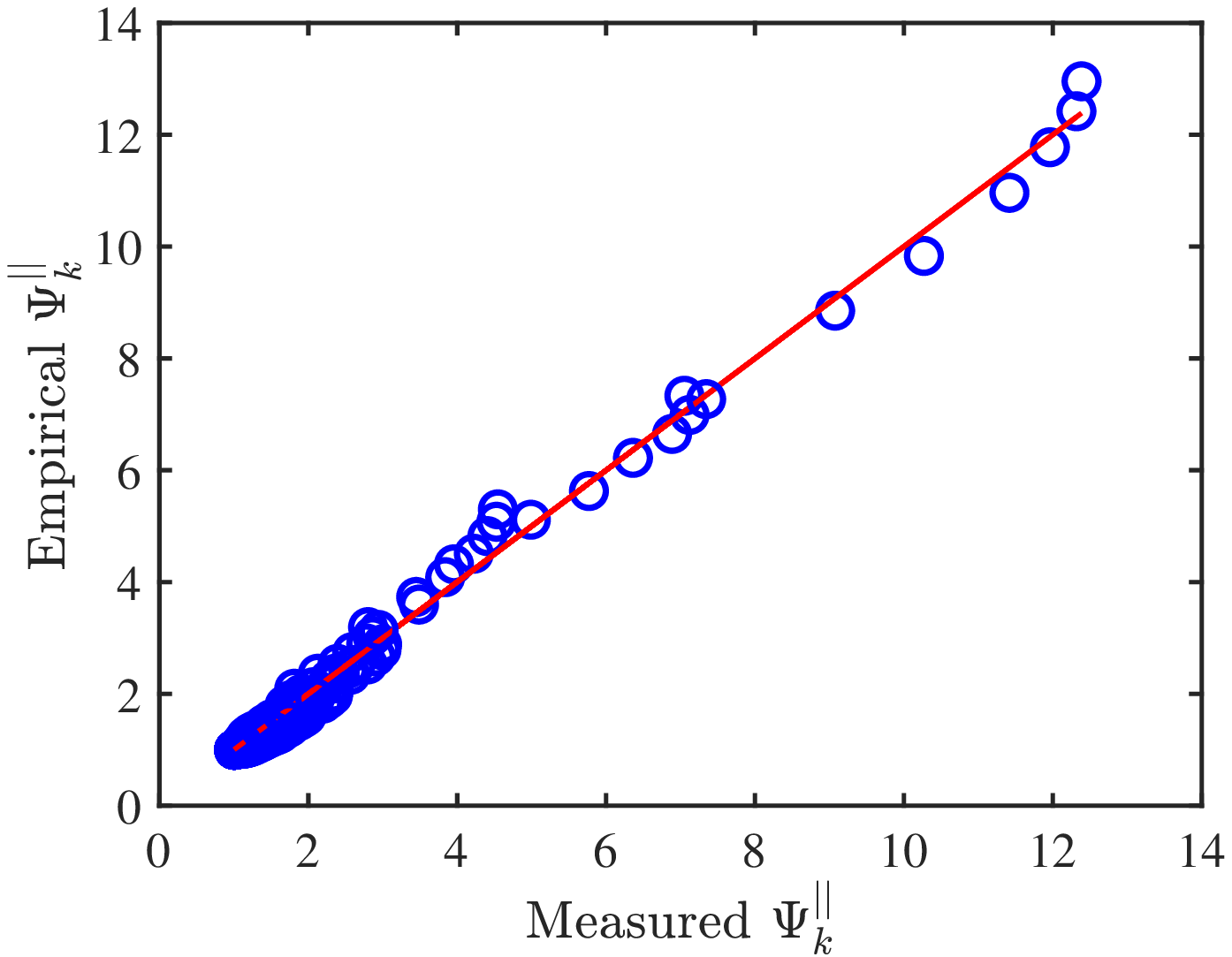}\hspace{0.04\textwidth}
    \includegraphics[scale=0.55]{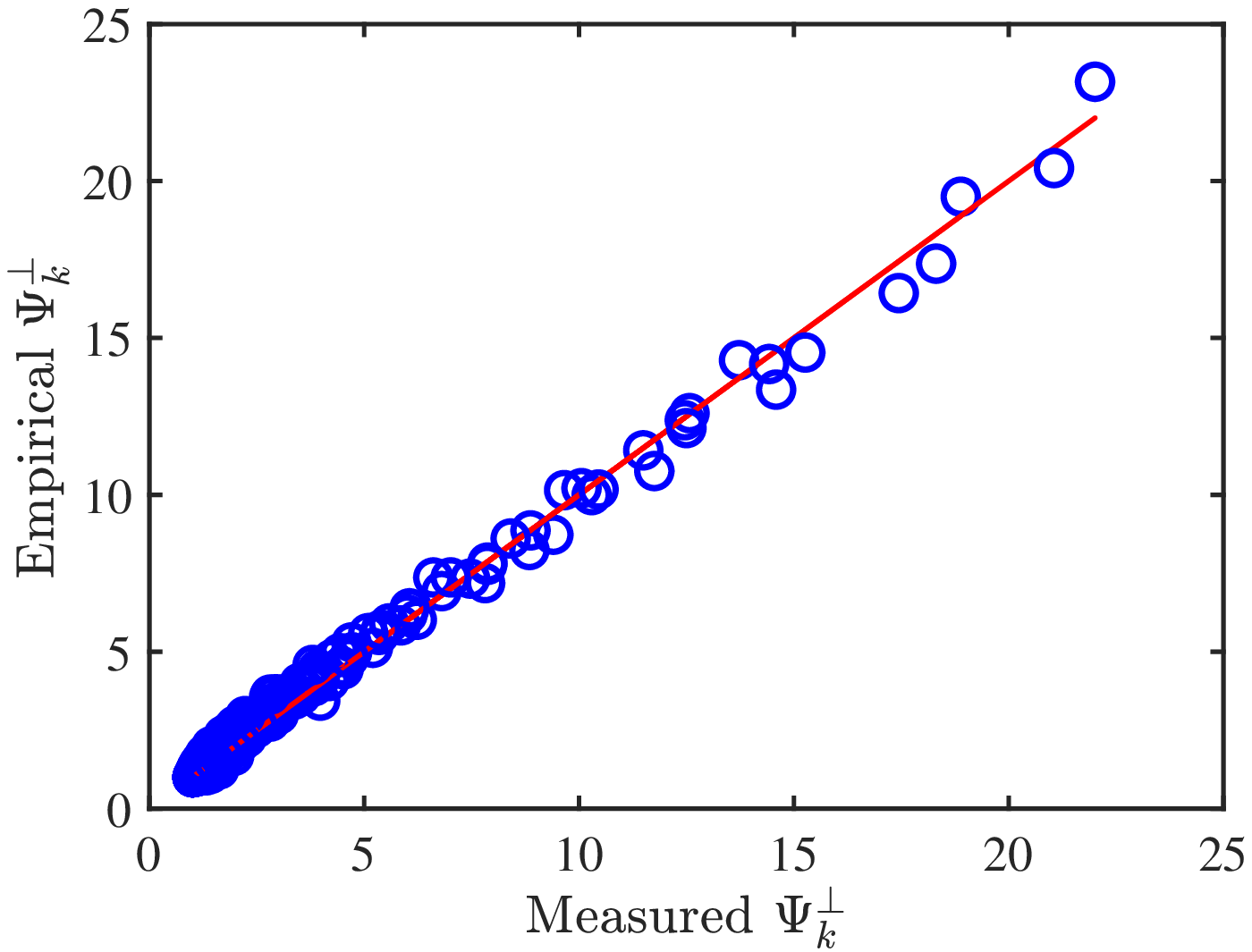}
    \caption{Shown are the predictions of Eqs. \ref{eq:psi}-\ref{eq:psi_cont3} and \ref{eq:psi_normal_cl_wall} for parallel forces, i.e., $i=1,3$ (left) and normal forces, i.e., $i=2$ (right) compared to the numerical measurements. }
    \label{fig:psi}
\end{figure}

Figure \ref{fig:psi} shows the prediction of $\Psi^{(i)}_k$ using the above equations for both parallel and normal forces. Larger values correspond to the computational cells with high aspect ratio or those situated closer to the wall. Ignoring wall effect on the geometric correction factor and letting $\Psi^{(i)}_k{=}1$ yields overprediction of the computational velocity of the cell as $u^{(i)}_c{\propto}(\Psi^{(i)}_kK^{(i)}_c)^{-1}$. As shown later, this over prediction becomes remarkable when particles travel very close to the wall which results in erroneous particle trajectory.

So far we considered the disturbance created by a small force that is applied to the center of a computational cell. This condition assumes that the particle force is applied only to a cell that contains the particle. However, in EL-PP simulations, this assumption does not necessarily hold, and the particle force is commonly distributed to the number of computational cells that are located within the stencil of the distribution function. Depending upon their distance from the force, they receive a fraction of this force and get disturbed differently. Now, in the next time step, when the fluid forces are to be computed, a function is similarly employed to interpolate the fluid quantities to the location of particle. During this process, the disturbance created in the surrounding computational cells in the previous time step will enter into the force calculations and depending on the stencil of this function, particle receives different disturbances. To accurately capturing the disturbance that particle receives, these effects must be accounted for in the correction scheme. \cite{esmaily2018} derived an analytical formulation for these effects for unbounded flows wherein the disturbance around the particle is symmetric. However, near a no-slip wall, the shape and strength of the disturbance field vary and it becomes more asymmetric. Below, we generalize the analytical expression of E\&H to account for the no-slip walls and a new analytical expression is derived. 

Suppose the particle force, $F^{(i)}_p$, is fed back to the background flow using a distribution function that has a certain bandwidth. Those computational cells that lie within the bandwidth receive a fraction of the force depending on their distance to the particle. Accordingly, the corresponding force that computational cell $j$ receives is expressed as,

\begin{equation}
    F^{(i)}_j = \beta_j F^{(i)}_p,
\end{equation}

\noindent where $\beta_j$ is the distribution coefficient (weight) corresponding to the computational cell $j$. When the particle forces (e.g., the drag that requires fluid velocity) are being calculated, the disturbance field is interpolated to the particle location from the neighbouring cells as,

\begin{equation}
    u^{(i)}_{c} = \sum_{j=1}^{nj}\gamma_j u^{(i)}_{c,j},
    \label{eq:uc_p}
\end{equation}

\noindent where $u^{(i)}_c$ is the disturbance that particle receives in $i$ direction and $\gamma_j$ is the interpolation coefficient corresponding to the computational cell $j$ that has computational velocity (disturbance velocity) of $u^{(i)}_{c,j}$. $n_j$ is the total number of adjacent computational cells that are employed for the interpolation. It is imperative to note that unlike staggered grids, in collocated arrangements, $\gamma_j$ and $\beta_k$ coefficients are direction independent. The question that arises here is how to compute the computational velocity of the adjacent computational cells, $u^{(i)}_{c,j}$, when they are imposed to a fraction of particle force. A naive way to obtain that, is to simply use Eq. \ref{eq:uc} for each cell with its given force, $F^{(i)}_j$, assuming that the computational cells are independent and only disturbed by their direct forces. In practice, however, this assumption does not hold and each computational cell gets disturbed not only by their direct force but also through the perturbations induced by the adjacent cells. For instance, when the computational cell $k$ is disturbed by its own force, $\beta_kF^{(i)}_p$, the created disturbance velocity in cell this cell pushes and perturbs the surrounding cells through $\alpha^{(i)}_{kj}$ that is the velocity ratio of cell $j$ generated by perturbation of cell $k$ to that of the computational cell $k$. This implies the fact that the disturbance created in computational cell, e.g., $j$, constitutes a combination of the one created by its own direct force and those created by the adjacent cells. Upon finding a closure for $\alpha^{(i)}_{kj}$, a linear superposition is valid if the created disturbance field meets the zero Reynolds number criterion. For unbounded flows and in the limit of zero Reynolds number, \cite{esmaily2018} showed that $\alpha^{(i)}_{kj}$ can be predicted using the Stokes solution that is the solution for the velocity field generated around a sphere slowly moving in an unbounded quiescent flow as

\begin{equation}
\alpha^{(i)}_{kj} = \frac{3}{4}r^{\prime -1}_{kj} \left(1+cos^2\theta^{(i)}_{kj}\right) + \frac{1}{4} r^{\prime -3}_{kj}\left(1-3cos^2\theta^{(i)}_{kj}\right),
\label{eq:stokes}
\end{equation}

\begin{figure}
    \centering
    \includegraphics[scale=0.3]{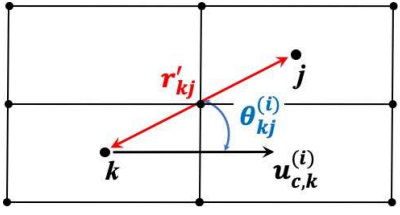}
    \caption{Schematic of computational cell $k$ that is disturbed by a small force and has disturbance velocity of $u^{(i)}_{c,k}$ which perturbs the adjacent computational cells through the modelled Stokes solution. $r^{\prime}_{kj}$ is the normalized distance between cell $k$ and $j$ with polar angle of $\theta^{(i)}_{kj}$ between the line passing through these cells and $i$ direction.}
    \label{fig:stokes}
\end{figure}

\noindent where $\theta^{(i)}_{kj}$ is the polar angle between the line passing through the computational cells $k$ and $j$ and the $i$ direction (Fig. \ref{fig:stokes}) and $r^{\prime}_{kj}$ is the distance between these two cells normalized by the characteristic length of the computational cell. The choice of this equation was inspired by the fact that the computational cell is treated as a solid object that moves in the fluid and consequently disturbs the surrounding fluid in a manner similar to a solid sphere. Using the prediction of this equation and a characteristic length of $0.28d_c$, they showed an excellent agreement with their numerical measurements. 

For the collocated grid arrangement used in this study, we found that Stokes solution (Eq. \ref{eq:stokes}) normalized with a smaller characteristic length of $0.25dc$ better predicts our numerical measurements. This was done by performing measurements similar to the previous parts. A small force in $i$ direction is applied to the computational cell $k$ located in the middle of a sufficiently large periodic box. At steady state, we measure the velocity of the perturbed cell ($k$) as well as those of its adjacent cells ($j$). The velocity ratio of these two cells, $u^{(i)}_{c,j}/u^{(i)}_{c,k}$, is $\alpha^{(i)}_{kj}$ by definition. For the sake of clarity, this parameter could be alternatively denoted by $b_{lmn}$ in which the subscript $lmn$ corresponds to the location of cell $j$, that is  $[la^{(1)},ma^{(2)},na^{(3)}]$ away from the computational cell $k$. As an example, $b_{100}$ represents the velocity ratio of cell $j$ to $k$ with $j$ being the immediate cell in the $(1)$ direction and right hand side of the perturbed cell $k$. Table \ref{tab:b_values} shows the prediction of Eq. \ref{eq:stokes} normalized with both $0.25d_c$ and $0.28d_c$ compared to our numerical measurements on the collocated grid arrangement for different aspect ratios. Better predictions are obtained by the former characteristic length. For the sake of comparison, we have also included the corresponding values of \cite{esmaily2018} that are based on the staggered grid arrangements, revealing a slight difference between these two arrangements. The difference is justified due to the fact that in collocated arrangements, unlike the face velocity, the cell-centered velocity is not necessarily divergence free, thereby causing small errors in the results compared to those of the staggered arrangements. 

\begin{table}[!h!]
\begin{center}
\def~{\hphantom{0}}
\begin{adjustbox}{width=\textwidth}
\begin{tabular} {lccccccccccc}
\hline
$a^{(2)}/a^{(1)}$ & $a^{(3)}/a^{(1)}$ & & & $b_{000}$ & $b_{100}$& $b_{010}$ & $b_{110}$ & $b_{001}$ & $b_{101}$ & $b_{011}$ & $b_{111}$ \\ 
\hline
1.0 & 1.0 & measured & collocated & 1.0 & 0.31 & 0.27 & 0.18 & 0.27 & 0.18 & 0.16 & 0.14\\
& & & staggered & 1.0 & 0.50 & 0.25 & 0.24 & 0.25 & 0.24 & 0.15 & 0.16\\
& & predicted & using $0.25d_c$ & 1.0 & 0.45 & 0.24 & 0.25 & 0.24 & 0.25 & 0.17 & 0.18\\
& & & using $0.28d_c$ & 1.0 & 0.50 & 0.27 & 0.27 & 0.27 & 0.27 & 0.19 & 0.20\\
&&&&&&&&&&&\\

1.0 & 2.0 & measured & collocated & 1.0 & 0.41 & 0.33 & 0.26 & 0.19 & 0.17 & 0.15 & 0.14\\
& & & Staggered & 1.0 & 0.62 & 0.33 & 0.34 & 0.17 & 0.18 & 0.13 & 0.15\\
& & predicted & using $0.25d_c$ & 1.0 & 0.56 & 0.31 & 0.31 & 0.15 & 0.16 & 0.13 & 0.14 \\
& & & using $0.28d_c$ & 1.0 & 0.61 & 0.35 & 0.34 & 0.17 & 0.18 & 0.15 & 0.16\\
&&&&&&&&&&&\\

2.0 & 4.0 & measured & collocated & 1.0 & 0.62 & 0.36 & 0.34 & 0.22 & 0.21 & 0.18 & 0.18\\
&  & & staggered & 1.0 & 0.83 & 0.31 & 0.34 & 0.16 & 0.17 & 0.13 & 0.13\\
& & predicted & using $0.25d_c$ & 1.0 & 0.81 & 0.24 & 0.25 & 0.12 & 0.12 & 0.10 & 0.11\\
& & & using $0.28d_c$ & 1.0 & 0.87 & 0.27 & 0.28 & 0.13 & 0.13 & 0.12 & 0.12\\
\hline
\end{tabular}
\end{adjustbox}
\caption{Measured $b_{lmn}$ values in comparison with the prediction of Eq. \ref{eq:stokes} normalized with the characteristic length of $0.25d_c$ and $0.28d_c$. Shown also includes the corresponding measured values from \protect\cite{esmaily2018} that are based on the staggered grid arrangement. }
\label{tab:b_values}
\end{center}
\end{table}

The next step is answering the question of how does $\alpha_{kj}$ change when the disturbance occurs close to a no-slip wall? One could substitute this parameter with the wall-bounded Stokes solution of a sphere moving in a quiescent flow and near a no-slip wall \citep{oneil1964,oneil1967}. Although there have been a few methods for simplifying such solution (e.g., \cite{chaoui2003}), it is expressed as expansions of spherical harmonics with the coefficients that are obtained iteratively as the solution of an infinite linear system. This makes the use of wall-bounded Stokes solution computationally expensive for EL approaches. 

An alternative remedy is the choice of the ``Stokeslet solution'' that is the flow field generated by a point force in a quiescent fluid. Direct analytical solutions are available for both unbounded and wall-bounded flows \citep{blake1971} that makes it more desirable and feasible to be implemented in EL approaches. Assuming that the ratio of the wall-bounded to the unbounded Stokes solution, $\alpha^{(i)}_{stk,b}/\alpha^{(i)}_{stk,un}$, approximately equals to the corresponding ratio of Stokeslet solution, $\alpha^{(i)}_{stkl,b}/\alpha^{(i)}_{stkl,un}$, an analytical expression for the wall adjustment to $\alpha^{(i)}_{kj}$ is derived (see Appendix B for the detailed Stokeslet solutions) as,

\begin{equation}
    \Phi^{(i)}_{kj} = \frac{(\alpha^{(i)}_{stkl,b})_{kj}}{(\alpha^{(i)}_{stkl,un})_{kj}} = 1 - \left[ \frac{ \frac{1}{|R_{kj}|} + \frac{(R^{(i)}_{kj})^2}{|R_{kj}|^3} +\frac{2x^{(2)}_kf^{(i)}_{kj}}{|R_{kj}|^6}} { \frac{1}{|r_{kj}|} + \frac{(r^{(i)}_{kj})^2}{|r_{kj}|^3} }\right],
    \label{eq:phi}
\end{equation}

\noindent where, 

\begin{equation}
    f^{(i)}_{kj} =  (-1)^i \left(x^{(2)}_k|R_{kj}|^3 - 3|R_{kj}|(R^{(i)}_{kj})^2x^{(2)}_k - |R_{kj}|^3R^{(2)}_{kj} + 3|R_{kj}|(R^{(i)}_{kj})^2R^{(2)}_{kj}\right)
    \label{eq:phi_cont1}    
\end{equation}

\begin{equation}
    r^{(i)}_{kj} = (x^{(i)}_j-x^{(i)}_k) \text{,} \quad |r_{kj}| = \sqrt{ \sum_{i=1}^{3}(r^{(i)}_{kj})^2}
    \label{eq:phi_cont2}
\end{equation}

\begin{equation}
    R^{(i)}_{kj} = 
    \begin{cases}
    r^{(i)}_{kj}, \quad i=1,3\\
    r^{(2)}_{kj}+2x^{(2)}_k, \quad i=2
    \end{cases}
    \text{,} \quad |R_{kj}| = \sqrt{ \sum_{i=1}^{3}(R^{(i)}_{kj})^2}
    \label{eq:phi_cont3}
\end{equation}

\noindent and $x^{(i)}_j$ and $x^{(i)}_k$ are the $i$ coordinate of the computational cell $j$ and $k$, respectively. Note that $\Phi^{(i)}_{kj}$ is not normalized by any characteristic length that makes it general for both staggered and collocated grid arrangements. It is imperative to mention that when the disturbance created by a particle is situated sufficiently away from the wall, both bounded and unbounded Stokeslet solutions become identical and this parameter becomes unity as

\begin{equation}
    \lim_{x^{(2)}_k \to \infty} \Phi^{(i)}_{kj} =1,
\end{equation}

\noindent which makes the model general for capturing the disturbance field created at any wall distance, a common scenario in wall-bounded particulate flows. Knowing the adjacent perturbations, now we can find the computational velocity of each cell and derive the analytical expression for $K^{(i)}_p$ as follows. 

For the particle force that is stationary and distributed to its neighbour cells, in the limit of steady state and zero Reynolds number, the computational velocity of cell $j$ is obtained as the superposition of disturbances created by its own force as well as its adjacent cells as expressed below 

\begin{equation}
    u^{(i)}_{c,j} = \sum_{k=1}^{n_k} \left[\frac{\alpha^{(i)}_{kj}\beta_{k}\Phi^{(i)}_{kj}}{\Psi^{(i)}_k}\right]\frac{-F^{(i)}_p}{3\pi \mu d_cK^{(i)}_c},
    \label{eq:u_cj}
\end{equation}

\noindent where $n_k$ is the total number of computational cells to which the particle force is distributed. In Eq. \ref{eq:u_cj} and what follows, no implicit summation over repeated indices is implied. Note that we keep the wall adjustment to the geometric correction factor, $\Psi^{(i)}_k$, in the bracket as it varies among the adjacent cells owing to their different wall-normal distances. Knowing the disturbance velocity for the computational cells around the particle, the disturbance velocity seen by the particle is obtained using Eqs. \ref{eq:uc_p} and \ref{eq:u_cj} as, 

\begin{equation}
    u^{(i)}_{c} = \sum_{j=1}^{n_j}\left[\gamma_j \sum_{k=1}^{n_k} \left[\frac{\alpha^{(i)}_{kj}\beta_{k}\Phi^{(i)}_{kj}}{\Psi^{(i)}_k}\right]\right]\frac{-F^{(i)}_p}{3\pi \mu d_cK^{(i)}_c},
\end{equation} 

\noindent where $n_j$ is the total number of computational cells from which the fluid properties are interpolated to the particle location. The analytical expression for $K^{(i)}_p$ is then derived as

\begin{equation}
    K^{(i)}_{p} = \sum_{j=1}^{n_j}\left[\gamma_j \sum_{k=1}^{n_k} \left[\frac{\alpha^{(i)}_{kj}\beta_{k}\Phi^{(i)}_{kj}}{\Psi^{(i)}_k}\right]\right].
    \label{eq:kp}
\end{equation}

\noindent In the limit of large wall distances, since both $\Psi^{(i)}_k$ and $\Phi^{(i)}_{kj}$ approach unity, the $K^{(i)}_p$ derived here becomes identical to that derived in E\&H. It is crucial to mention that with this formulation all wall adjustments have been accounted for in the derivation of $K^{(i)}_p$. 

For cases where only ``box filtering" (zeroth order) is utilized, i.e., the particle only disturbs one cell from which the fluid properties are interpolated to the particle too ($n_k{=}n_j{=}1$), we have $\gamma_j{=}\beta_k{=}\alpha^{(i)}_{kj}{=}\Phi^{(i)}_{kj}{=}1$. In this case, $K^{(i)}_{p}{=}1/\Psi^{(i)}_{k}$, wherein subscript $k$ corresponds to the cell in which the particle lies. In such a simple case, $K^{(i)}_{p}$ becomes only the wall effect on the correction scheme. 

%%%%%%%%%%%%%%%%%%%%%%%%%%%%%%%%%%%%%%%%%%%%%%%%%%%%%
\subsection{Correction for the finite Reynolds number}
%%%%%%%%%%%%%%%%%%%%%%%%%%%%%%%%%%%%%%%%%%%%%%%%%%%%%
The Stokes drag used in Eq. \ref{eq:uc} is only valid for disturbances created with zero Reynolds number. To account for the higher Reynolds number effects, a Schiller-Naumann correction factor, analogous to the finite Reynolds number adjustment to the Stokes drag of a sphere \citep{clift}, 

\begin{equation}
    C_r = 1 + 0.15Re^{0.687}_c
    \label{eq:cr}
\end{equation}

\noindent can be used \citep{esmaily2018}; where, $Re_c{=}u_cd_c/\nu$ is defined as the Reynolds number of the computational cell based on its velocity and diameter. A wall-modified version of this equation has been empirically derived by \cite{zeng2009}, yet our results show that the use of Schiller-Naumann expression (Eq. \ref{eq:cr}) still yields better predictions for the studied wall-bounded cases. This expression captures only the change to the Stokes drag for higher $Re_c$ cases, however, the complexity of the asymmetric disturbance field at high $Re_c$ breaks the use of Eq. \ref{eq:stokes}, and the linear superposition employed in the derivation of Eq. \ref{eq:kp} does not hold anymore. Therefore, it is argued that for high $Re_c$, a more elaborate formulation might be required. As explained later, our results illustrate that the current formulation generates reasonable results for cases with $Re_c$ of up to 10. For larger $Re_c$, \cite{balachandar2019} showed that the need for the correction diminishes owing to the fact that the particle with larger $Re_p$ does not stay in its own disturbance, and in the next time step, it sees a more undisturbed flow for the force calculations. Although this effect is partly captured by introducing a temporal correction factor for finite exposure time, $C^{(i)}_t$, explained in the next part, a comprehensive study on the necessity of the correction scheme for a range of particle Reynolds number is left for future investigations.  

%%%%%%%%%%%%%%%%%%%%%%%%%%%%%%%%%%%%%%%%%%%%%%%%%%%%
\subsection{Correction for the finite exposure time}
%%%%%%%%%%%%%%%%%%%%%%%%%%%%%%%%%%%%%%%%%%%%%%%%%%%%
A particle moving in the computational domain spends a limited time within each computational cell and disturbs the cell for a finite time. This finite time exposure of particle has to be accounted for in Eq. \ref{eq:uc}, separately. The unsteady term in this equation is considered for the unsteady effect of a stationary force and does not include its limited exposure time. To demonstrate the need for this correction factor, consider a high velocity particle whose residency time within the computational cell with diameter of $d_c$ is $d_c/u_p$, which is much smaller than the response time of the fluid to the particle force, $d^2_c/\nu$, i.e., 

\begin{equation}
    \frac{d_c}{u_p} \ll \frac{d^2_c}{\nu}% \longrightarrow Re_c\gg1
\end{equation}

If particle size is assumed to be in the same order of the computational cell, i.e., $d_p{\sim}d_c$, then this criterion results in $Re_p{\gg}1$. In such scenario, particle passes through the grid quickly with negligible disturbance that obviates any need for the correction. Conversely, for particles with $Re_p{\ll}1$, their large exposure time allows them to sufficiently perturb the computational cell which underscores the need for the correction. This effect should be accounted for in Eq. \ref{eq:uc} separately as for cases with $Re_p{\gg}1$, this equation yields erroneously large computational velocity, which conceptually should be zero. In order to account for this effect, one could track the particle within each computational cell and only integrate Eq. \ref{eq:uc} over the period of time that particle spends in the cell and upon its exit the force becomes zero. To avoid the complexity added by this, we use the corresponding correction factor of E\&H as, 

\begin{equation}
    C^{(i)}_t = 1 - \frac{\tau^{(i)}_c}{\Delta t^{(i)}} \left( 1- \exp\left( -\frac{\Delta t^{(i)}}{\tau^{(i)}_c}\right) \right),
    \label{eq:ct}
\end{equation}

\noindent where, 

\begin{equation}
    \Delta t^{(i)} = \frac{a^{(i)}}{|u^{(i)}_p|} \quad \text{and} \quad \tau^{(i)}_c = \frac{d^2_c}{12\nu K^{(i)}_c},
    \label{eq:tc_delta_t}
\end{equation}

\noindent where $\tau^{(i)}_c$ and $\Delta t^{(i)}$ are respectively the computational cell relaxation time and the particle residence time in $i$ direction of the computational cell, respectively. The factor $C^{(i)}_t$ is a time-average of the solution of Eq. \ref{eq:uc} for a small force that is applied on top of a computational cell. Accordingly, for a particle with $Re_p{\gg}1$, its exposure time to the cell becomes large, $\Delta t{\rightarrow}0$ and using Eq. \ref{eq:ct}, $C^{(i)}_t{\rightarrow}0$ which eliminates any need for correction. However, for slow particles ($Re_p{\ll}1$), $\Delta t{\rightarrow}\infty$ and $C^{(i)}_t{\rightarrow}1$ which enforces the correction. In the next part, we combine all these correction factors and explain the steps in order to correct the PP approach. 

%%%%%%%%%%%%%%%%%%%%%%%%%%%%%%%%%%%%%
\subsection{The correction algorithm}
%%%%%%%%%%%%%%%%%%%%%%%%%%%%%%%%%%%%%
The entire correction scheme reduces to the computation of Eq. \ref{eq:uc} that is solved concurrently with the equation of motion of the particle (Eq. \ref{eq:newton_2ndlaw}). Although one could simply use any time integration scheme for these two equations, we use an explicit method for the results presented in this work. Therefore, knowing the $u^{(i)}_c$ and $u^{(i)}_p$ from previous time step, the following procedure is used. 

\begin{enumerate}
 
\item Compute the disturbed velocity at the location of particle, $u^{(i)}_d$, that is readily available in the standard PP packages.
\item Compute the undisturbed velocity at the location of particle, $u^{(i)}_f$, by using Eq. \ref{eq:uf} and having the computational velocity at the location of particle, $u^{(i)}_c$.
\item Compute the total fluid force exerted at the location of particle, $F^{(i)}$.
\item Update the velocity of particle, $u^{(i)}_p$, using Eq. \ref{eq:newton_2ndlaw}.
\item Calculate $K^{(i)}_c$ using Eqs. \ref{eq:kc} and \ref{eq:kc_cont} based on the grid size [$a^{(1)},a^{(2)},a^{(3)}$] in which particle is located.
\item Identify the location of surrounding cells to which the particle force is distributed ($n_k$). 
\item Identify the location of surrounding cells from which the fluid quantities are interpolated to the location of particle ($n_j$).
\item From the location of particle to the above computational cells, calculate $r^{'}_{kj}$ and $\theta^{(i)}_{kj}$ and thereby $\alpha^{(i)}_{kj}$ using Eq. \ref{eq:stokes}.
\item In the presence of no-slip walls, calculate $\Phi^{(i)}_{kj}$ and $\Psi^{(i)}_{k}$ based on Eqs. \ref{eq:psi}-\ref{eq:psi_cont3} and Eqs. \ref{eq:phi}-\ref{eq:phi_cont3}, respectively. 
\item Compute $K^{(i)}_p$, using Eq. \ref{eq:kp} and knowing $\beta_k$, $\gamma_j$, $\alpha^{(i)}_{kj}$, $\Phi^{(i)}_{kj}$ and $\Psi^{(i)}_{k}$.
\item Compute $Re_c$ and thereby $C_r$ using Eq. \ref{eq:cr}.
\item Compute $\tau^{(i)}_c$ and $\Delta t^{(i)}$ using Eq. \ref{eq:tc_delta_t} and thereby $C^{(i)}_t$ using Eq. \ref{eq:ct}.
\item Compute $K^{(i)}_t$ using Eq. \ref{eq:Kt} by knowing $K^{(i)}_c$, $K^{(i)}_p$, $C_r$ and $C^{(i)}_t$. 
\item Update $u^{(i)}_c$ using Eq. \ref{eq:uc}. 
 
\end{enumerate}

The initial condition for the procedure above is $u^{(i)}_c{=}0$ corresponding to the undisturbed fluid phase before injecting particles. For isotropic grids, the simplified formulation introduced in Appendix A could be used to compute $\Psi^{(i)}_{k}$ in the step 9 above. It is imperative to mention that for particle-laden flows wherein the particle time scale is smaller than that of the fluid, sub-cycling for particles' motion is typically performed. Particles are advanced during the frozen flow time scale and then at the end of the sub-cycling their force will be applied to the background flow. For such cases, the correction should be enforced once the sub-cycling is finished as that is when the flow is altered by the presence of particles. In the next section, the results of the present correction scheme are discussed and the accuracy of the scheme is assessed. 
%%%%%%%%%%%%%%%%%
\section{Results}
%%%%%%%%%%%%%%%%%
\label{sec:results}
In this section, the present correction scheme is verified by performing several test cases involving unbounded and wall-bounded flows. Different flow parameters and grid aspect ratios are carried out in order to assess the generality and robustness of the model for a wide range of applications. In the first set of computations, we start with settling velocity of a particle in an unbounded flow wherein the wall effects do not appear and the model for the collocated arrangements is validated against the analytical solution. In the second set of test cases, the model is validated for velocity of a particle settling parallel and close to a no-slip wall. Test cases at different wall distances, ranging from near to sufficiently away from the wall, are performed to test the model for possible situations that happen in particle-laden flows. Different grid aspect ratios representative of typical turbulent channel flows are used in these tests. In the third set of assessments, the model will be employed to freely falling motion of a particle normal to the wall. The grid resolution for all cases was set to be $128^3$ as it was found to be sufficient to produce the results that are grid independent.

The three shared non-dimensional flow parameters among cases are those defined based on the Stokes flow in an unbounded configuration. The first one is the Stokes parameter, $St$, defined as the ratio of the particle relaxation time, $\tau_p$, to the fluid time scale, $\tau_f$, as, 

\begin{equation}
    St=\frac{\tau_p}{\tau_f},
    \label{eq:st}
\end{equation}

\noindent where, 

\begin{equation}
    \tau_p = \frac{\rho_pd^2_p}{18\mu},
\end{equation}

\noindent and,

\begin{equation}
    \tau_f = \frac{\min\left(a^{(i)}\right)^2}{\nu},
\end{equation}

The second parameter is the particle Reynolds number as,

\begin{equation}
    Re^{Stk}_p = \frac{|\mathbf{u}^{Stk}_{s}|d_p}{\nu},
     \label{eq:Rep_Stk}
\end{equation}

\noindent where, 

\begin{equation}
    \mathbf{u}^{Stk}_{s}=\left(1-\rho_f/\rho_p \right)\tau_p\mathbf{g},
    \label{eq:u_set_stokes}
\end{equation}

\noindent is the particle settling velocity under gravity, $\mathbf{g}$, in an unbounded Stokes flow with $\rho_f$ and $\rho_p$ being the fluid and particle densities, respectively. The third parameter that has three components as the ratio of particle diameter to the grid size is 

\begin{equation}
    \Lambda^{(i)} = \frac{d_p}{a^{(i)}}.
\end{equation}

\begin{figure}
    \centering
    \includegraphics[scale=0.32]{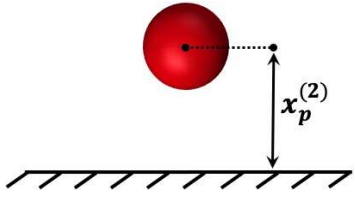}
    \caption{Particle located at wall distance of $x^{(2)}_p$ near a no-slip wall.}
    \label{fig:particle_near_wall}
\end{figure}

For wall-bounded test cases, another non-dimensional parameter that is the normalized wall distance from the bottom of particle is defined as (see Fig. \ref{fig:particle_near_wall}) 

\begin{equation}
    \delta_p = \frac{x^{(2)}_p}{d_p}-0.5,
    \label{eq:delta_p}
\end{equation}

\noindent wherein $x^{(2)}_p$ is the wall distance from the center of particle. It should be noted that in wall-bounded cases, since the particle drag coefficient changes due to the presence of wall, its actual Reynolds number then differs from its unbounded counterpart expressed by Eq. \ref{eq:Rep_Stk}.  

For the first and second test cases, we evaluate the accuracy of the model based on the errors in the settling, drifting and total velocities of the particle compared to their reference values. Accordingly, the particle velocity, $\mathbf{u}_p(t)$, is decomposed into two components; parallel and perpendicular to the reference velocity of $\mathbf{u}_r$. The parallel component is expressed as,

\begin{equation}
    \mathbf{u}^{||}_p = \frac{\mathbf{u}_r \cdot \mathbf{u}_p(t)}{|\mathbf{u}_r|^2}\mathbf{u}_r,
    \label{eq:up_parallel}
\end{equation}

\noindent while the perpendicular component is obtained by, 

\begin{equation}
    \mathbf{u}^{\perp}_p = \mathbf{u}_p(t) - \mathbf{u}^{||}_p.
    \label{eq:up_perp}
\end{equation}

The errors in these two velocity components are then calculated based on the following metrics,

\begin{equation}
e^{\parallel} = \frac{\overline{\mathbf{u}^{\parallel}_p(t).\mathbf{u}_r}}{|\mathbf{u}_r|^2} -1;
\label{eqn:paral_error}
\end{equation}

\begin{equation}
e^{\perp} = \frac{\overline{|\mathbf{u}^{\perp}_p(t)|}}{|\mathbf{u}_r|},
\label{eqn:perp_error}
\end{equation}

\noindent where, overbar $\overline{()}$ denotes the time averaging. Finally, error in the total velocity compared to the reference velocity is obtained as, 

\begin{equation}
e = \frac{\overline{|\mathbf{u}_p(t)-\mathbf{u}_r|}}{|\mathbf{u}_r|}.
\end{equation}

The reference velocity, $\mathbf{u}_r$, is the settling velocity of particle that is defined differently for each case depending on the corresponding drag coefficient. 

%%%%%%%%%%%%%%%%%%%%%%%%%%%%%%%%%%%%%%%%%%%%%%%%%%
\subsection{Settling particle in an unbounded flow}
%%%%%%%%%%%%%%%%%%%%%%%%%%%%%%%%%%%%%%%%%%%%%%%%%
In the test cases here, we first validate the present correction scheme for the unbounded flows in order to assess the new presented $K^{(i)}_c$ equation and the new characteristic length employed for normalization of Eq. \ref{eq:stokes}. Settling velocity of a particle in an unbounded periodic domain is performed. For the results of this part, we neglect the wall effects by setting $\Psi^{(i)}_k{=}\Phi^{(i)}_{kj}{=}1$. For all test cases, a particle that is initially stationary, $u^{(i)}_p{=}0$, thus $u^{(i)}_c{=}0$, and located in an unbounded flow settles under gravity and in the presence of the stokes drag force. Following the advice by \cite{horwitz2016}, gravity vector is chosen as $\mathbf{g}{=}(1,(1+\sqrt{5})/2,\exp(1))/|\mathbf{g}|$ so that particle sweeps through different locations among its adjacent computational cells ensuring that the model is capable of handling any arbitrary positioning of particle. The particle equation of motion in a quiescent fluid is then written as    

\begin{equation}
\frac{d\mathbf{u}_p}{dt} = \left(1-\frac{\rho_f}{\rho_p} \right)\mathbf{g} - \frac{f}{\tau_p} \mathbf{u}_p
\label{eq:dup_dt}
\end{equation}

\noindent where $f$ corresponds to any adjustment factor to the Stokes drag coefficient that is unity for the studied cases in this part. Accordingly, the analytical solution for the particle velocity for Stokes flow is obtained as 

\begin{equation}
    \mathbf{u}^{Stk}(t) = \mathbf{u}^{Stk}_{s} \left(1 - \exp(-\frac{t}{\tau_p})\right)
    \label{eq:u_ref_unb}
\end{equation}

\noindent where $\mathbf{u}^{Stk}_{s}$ is the settling velocity as provided in Eq. \ref{eq:u_set_stokes} and serves as the reference velocity. Table \ref{tab:error_unbounded} shows six different cases with various flow parameters and grid aspect ratios for all which the error in settling velocity of the particle without the correction is remarkably large. Errors in settling, drifting and total velocities of the particle predicted with and without the present correction scheme are compared. Additionally, the corresponding values from E\&H are listed for comparison. In general, the present scheme reduces the errors with the same order of magnitude as E\&H, however, for cases with large size particles such as case U02, the embedded error in the collocated arrangement that appears in the computation of Eq. \ref{eq:stokes} inevitably yields larger values compared to the staggered arrangement. It is worth mentioning that the time step used for the computations of the current cases is half of those reported in \cite{horwitz2016} so that the Peclet number of $Pe{=}6\nu_f\Delta t/\min(a^{(i)})^2{=}0.18$ as well as particle Courant number of $CFL_p{=}\Delta t/\tau_p{=}0.003$ are satisfied. 

\begin{figure}
    \centering
    \includegraphics[scale=0.65]{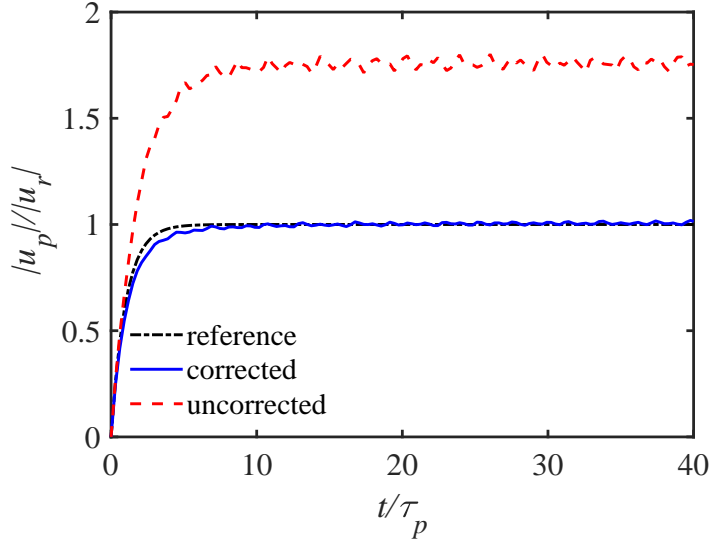}
    \caption{Plotted is the velocity of a settling particle as a function of time in an unbounded domain. Analytical solution (dash-dotted black), prediction of the present correction scheme (solid blue) as well as the uncorrected scheme (dashed red) are compared. The reference velocity, $u_r$, used for normalization is the particle settling velocity in Stokes flow given by Eq. \ref{eq:u_set_stokes}. Results pertain to case U01 from Tab. \ref{tab:error_unbounded}. }
    \label{fig:U01}
\end{figure}

Figure \ref{fig:U01} shows the particle velocity of case U01 as a function of time with and without the correction scheme. As illustrated, the present correction scheme produces excellent result compared to the reference velocity.  

\begin{table}
\begin{center}
\def~{\hphantom{0}}
\begin{adjustbox}{width=\textwidth}
\begin{tabular} {lcccccccc}
\hline
Case & $Re^{Stk}_p$ & $St$ &  $\Lambda^{(1)}$ & $\Lambda^{(2)}$ & $\Lambda^{(3)}$  &  \thead{uncorrected \\ $e^{\parallel}$ \quad \quad $e^{\perp}$ \quad \quad $e$} & \thead{E\&H\\ $e^{\parallel}$ \quad \quad $e^{\perp}$ \quad \quad $e$} & \thead{present model\\ $e^{\parallel}$ \quad \quad $e^{\perp}$ \quad \quad $e$} \\ 
\hline
U01 & 0.1 & 10.0 & 1.0 & 1.0 & 1.0 & 78.94 \quad 0.074 \quad 78.94 & 0.83 \quad 0.44 \quad 1.00 & 0.59 \quad 0.74 \quad 1.05 \\ 
U02 & 0.1 & 10.0 & 5.0 & 5.0 & 5.0 & 392.14 \quad 0.25 \quad 392.14 & 1.70 \quad 5.20 \quad 7.50 & -1.98 \quad 7.40 \quad 10.97\\
U03 & 0.1 & 10.0 & 5.0 & 0.5 & 0.5 &  57.40 \quad 7.98 \quad 57.96 & -2.00 \quad 1.80 \quad 2.90 & -3.91 \quad 2.07 \quad 4.59\\
U04 & 0.1 & 10.0 & 4.0 & 2.0 & 0.2 &  51.22 \quad 10.65 \quad 51.32 & -3.50 \quad 6.00 \quad 7.30 & -4.62 \quad 2.61 \quad 5.70\\
U05 & 0.5 & 10.0 & 1.0 & 1.0 & 1.0 & 68.64 \quad 0.08 \quad 68.64 & 4.30 \quad 2.00 \quad 4.70 & 4.83 \quad 2.29 \quad 5.35\\
U06 & 0.1 & 0.25 & 1.0 & 1.0 & 1.0 & 78.73 \quad 0.57 \quad 78.73 & 0.43 \quad 0.86 \quad 1.40 & -0.1 \quad 1.86 \quad 2.85\\
\hline
\end{tabular}
\end{adjustbox}
\caption{Listed are the percentage errors for settling, drifting and total velocity of a particle settling in an unbounded domain. Results with and without the present correction scheme are compared with the corresponding values from E\&H. Various cases with different particle diameter to gird sizes, $\Lambda$, particle Reynolds numbers, $Re_p$, and particle Stokes numbers, $St$, are shown for validation. }
\label{tab:error_unbounded}
\end{center}
\end{table}

%%%%%%%%%%%%%%%%%%%%%%%%%%%%%%%%%%%%%%%%%%%%%%%%%%%
\subsection{Settling particle parallel to the wall}
%%%%%%%%%%%%%%%%%%%%%%%%%%%%%%%%%%%%%%%%%%%%%%%%%%%
As the first step toward validating wall effects in the present correction scheme, velocity of a particle settling parallel to a no-slip wall is tested at different wall distances. In order to illustrate the need for the present scheme, results with and without accounting for $\Psi^{(i)}_k$ and $\Phi^{(i)}_{kj}$ in the formulation are compared against the reference. As listed in Tab. \ref{tab:error_parallel}, different flow parameters, grid aspect ratios and particle to grid sizes are carried out to assess the capability of the model for a wide range of applications. Similar to the preceding section, the errors in settling, drifting and total velocities of the particle are measured and compared among different schemes. For the studied cases, a particle that is initially located at a normalized wall gap, $\delta_p$, released to reach its settling velocity under a gravity vector of $\mathbf{g}{=}(\exp(1),0,(1+\sqrt{5})/2)/|\mathbf{g}|$ that guarantees the particle's motion on a plane parallel to the wall. In reality, the particle experiences a lateral force \citep{vasseur1977,takemura2003}, yet in this study other directions are neglected in order to isolate the parallel motion. The particle's equation of motion in the presence of wall follows Eq. \ref{eq:dup_dt} with the correction factor of $f$ that is employed based on the work of \cite{zeng2009}. In their work, an empirical drag coefficient is derived as a function of normalized wall gap, $\delta_p$, and the relative Reynolds number, $Re_p$, for a spherical object moving parallel to the wall and in a quiescent flow as, 

\begin{equation}
    C^{w,||}_d = \frac{24}{Re_p}f^{||}(\delta_p,Re_p),
    \label{eq:cd_wall_paral}
\end{equation}

\noindent where $f^{||}(\delta_p,Re_p)$ is the correction factor to the Stokes drag including two terms as

\begin{equation}
    f^{||}(\delta_p,Re_p) = f^{||}_1(\delta_p)f^{||}_2(\delta_p,Re_p),
    \label{eq:f_parallel}
\end{equation}

\noindent where, 

\begin{equation}
    f^{||}_1(\delta_p) = \left[1.028 - \frac{0.07}{1+4\delta^2_p} - \frac{8}{15}\log \left(\frac{270\delta_p}{135+256\delta_p}\right)\right],
\end{equation}

\begin{equation}
    f^{||}_2(\delta_p,Re_p) = \left[ 1+0.15\left(1-\exp\left(-\sqrt{\delta_p}\right)\right)Re_p^{\left(0.687+0.313\exp\left(-2\sqrt{\delta_p}\right)\right)}\right].
\end{equation}

$f^{||}_1(\delta_p)$ captures the wall effects on the Stokes drag for zero $Re_p$ that becomes unity for large $\delta_p$, recovering the Stokes drag coefficient. $f^{||}_2(\delta_p,Re_p)$, however, handles the wall-modified finite Reynolds number effects to the Stokes drag that converts to the Standard Schiller-Naumman correction factor \citep{clift} when particle travels sufficiently away from the wall. 

For the first cases studied in this part, $Re_p$ is very small, thus only $f^{||}_1(\delta_p)$ holds and the particle velocity is directly solved as, 

\begin{equation}
    \mathbf{u}^{w,||}(t) = \mathbf{u}^{w,||}_s \left(1 - \exp\left(-\frac{t}{\tau_p}f^{||}_1(\delta_p)\right)\right)
    \label{eq:up_reference_parallel}
\end{equation}

\noindent where, $\mathbf{u}^{w,||}_s$ is the particle settling velocity in parallel motion to the wall and in the limit of $Re_p{\sim}0$ as, 

\begin{equation}
    \mathbf{u}^{w,||}_s = \left(1-\frac{\rho_f}{\rho_p} \right)\frac{\tau_p\mathbf{g}}{f^{||}_1(\delta_p)}
    \label{eq:u_set_paral}
\end{equation}

Based on this drag formulation, the actual particle relaxation time in the presence of wall then becomes,

\begin{equation}
    \tau^{w,||}_p = \frac{\tau_p}{f^{||}_1(\delta_p)}
\end{equation}

Results based on the prediction of different schemes are compared with the reference given by Eq. \ref{eq:u_set_paral}. Following the metrics presented in the preceding section, the errors in settling, drifting and total velocities are measured. Table \ref{tab:error_parallel} shows these errors for the studied cases of this part which includes five sets, each of which has six cases corresponding to settling at different normalized wall gaps. Results with and without the wall correction factors on the correction scheme, $\Psi^{(i)}_k$ and $\Phi^{(i)}_{kj}$, are compared together with those of the uncorrected scheme to quantify the need for the wall-modified corrections scheme. For all the sets studied in this part, the particle Reynolds number of $Re^{Stk}_p{=}0.1$ and Stokes number of $St{=}10$ that are based on unbounded parameters, are kept constant. In practice, however, the actual particle Reynolds number decreases when it gets closer to the wall owing to the larger drag and this effect is studied separately in the next part. 

Sets A and B correspond to isotropic grid configuration with two different particle diameter to grid sizes, whereas the rest, C-F, pertain to anisotropic grids with various aspect ratios. The grid resolution used in the latter are those commonly encountered in the turbulent channel flows. The first observation from Tab. \ref{tab:error_parallel} is that the errors for the uncorrected scheme is significantly large for all cases, necessitating the need for correcting the Point-Particle approach even in the presence of a no-slip wall. In addition, consistent with observation of E\&H, the error in uncorrected results increases proportional to $(\Lambda^{(1)}\Lambda^{(2)}\Lambda^{(3)})^{1/3}{\propto}d_p/d_c$. As an example, the error in total velocity of the uncorrected scheme for case C1 is two order of magnitude smaller than that of case B1 wherein the volume ratio of particle to the grid is much greater. 

In the first place, one could correct the PP results with the unbounded version of the present correction scheme wherein wall effects are ignored, i.e., $\Psi^{(i)}_k{=}\Phi^{(i)}_{kj}{=}1$. As listed in Tab. \ref{tab:error_parallel}, for wall distances very close to the wall, such as $\delta_p{=}0.05$ and $0.5$, the unbounded version under predicts the particle velocity with negative errors on the same order of magnitude as the uncorrected scheme. These large errors in the near wall results are due to the overprediction in the computations of the disturbance velocity of the unbounded correction scheme, while particle in practice receives much smaller $u_c$ from the background flow near the no-slip boundary. When particle gets away from the wall, however, the predicted disturbance field using unbounded version becomes more accurate and reduces the errors significantly (see cases at $\delta_p{=}\infty$).
%In reality, the shape of the disturbance field becomes asymmetric near the wall and it decays faster toward the wall to satisfy the no-slip boundary condition to the wall, that is not accounted for in the unbounded version of the present correction scheme. 

\begin{figure}
    \centering
    \includegraphics[scale=0.55]{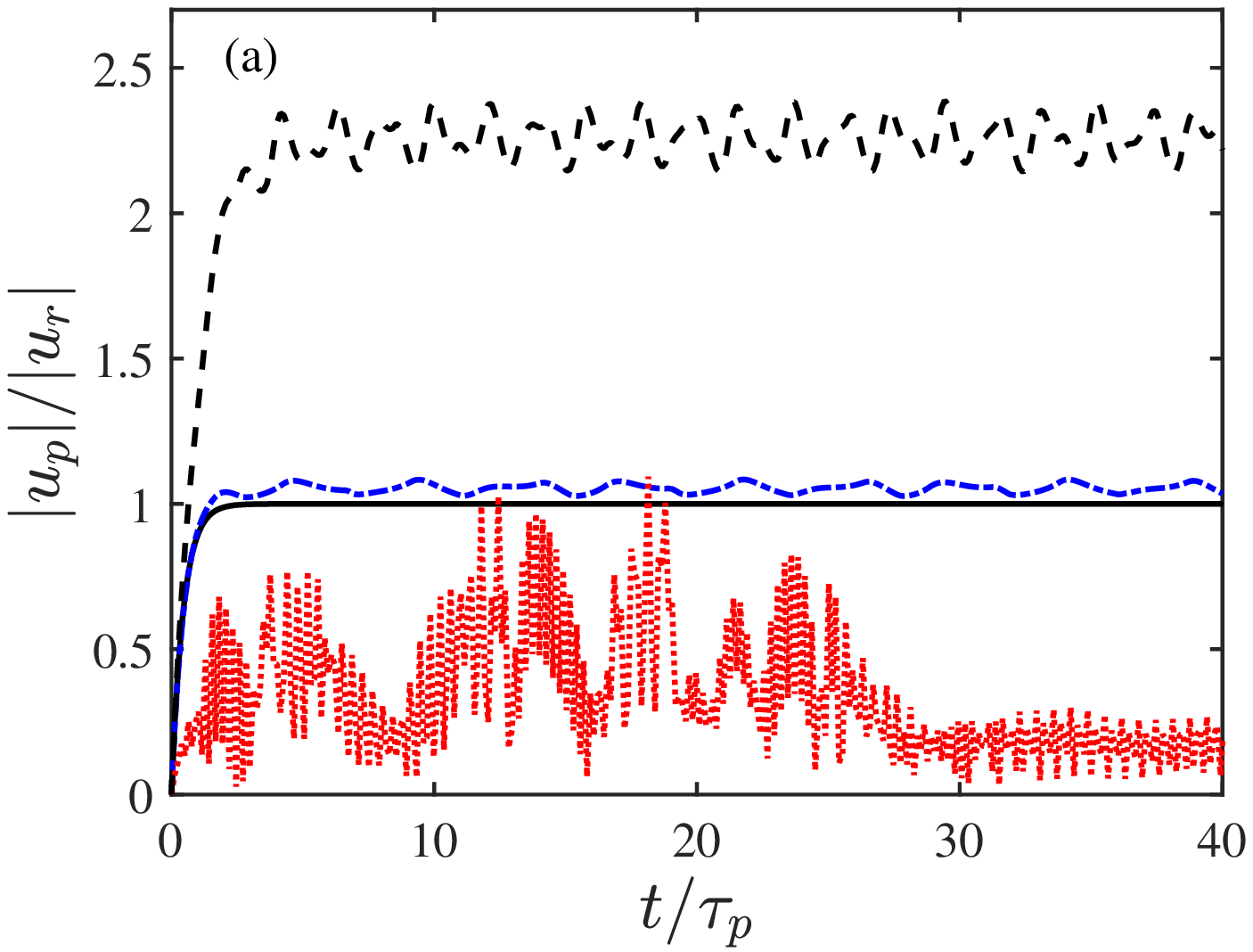}\hspace{0.04\textwidth}
    \includegraphics[scale=0.55]{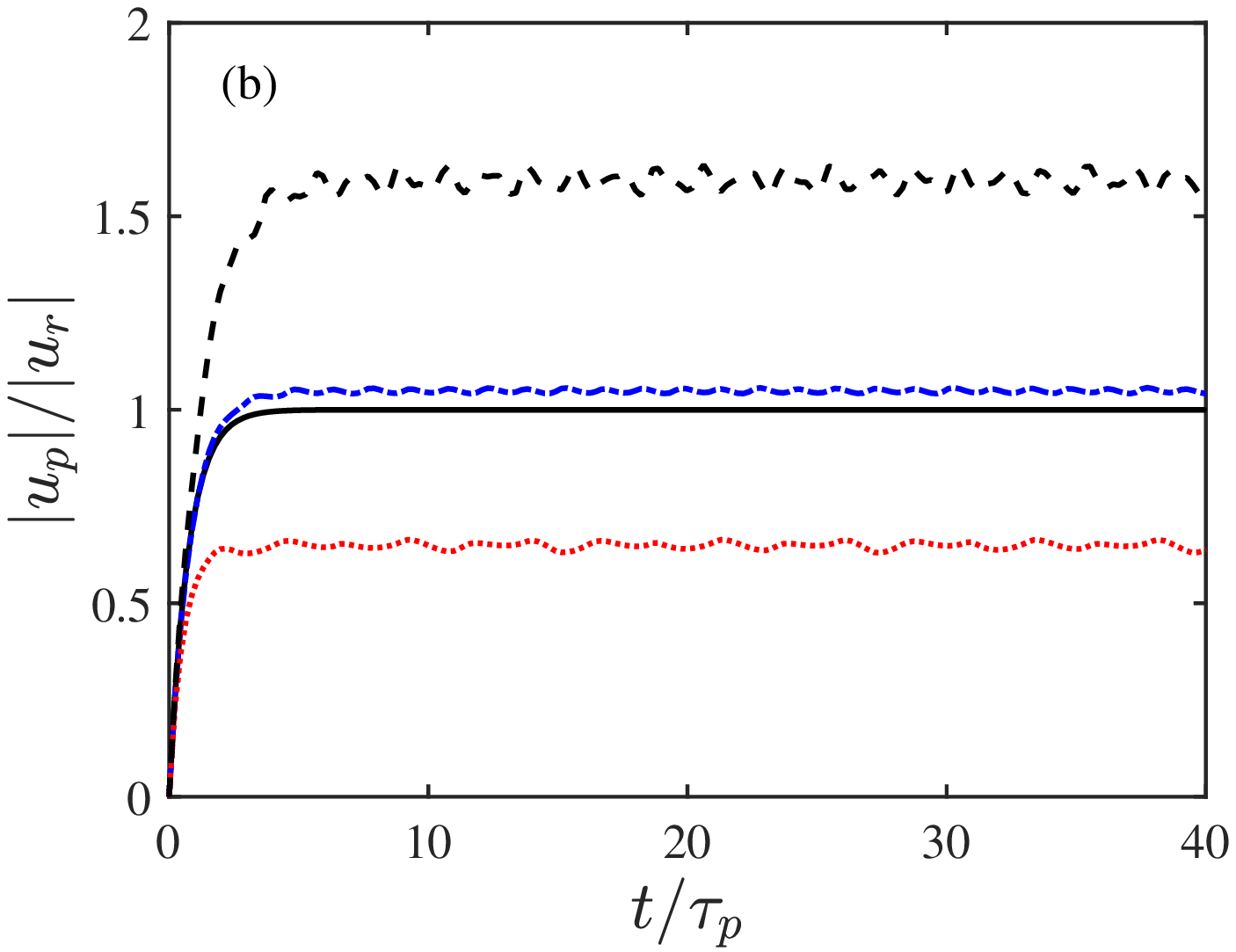}\vspace{0.04\textwidth}
    \includegraphics[scale=0.55]{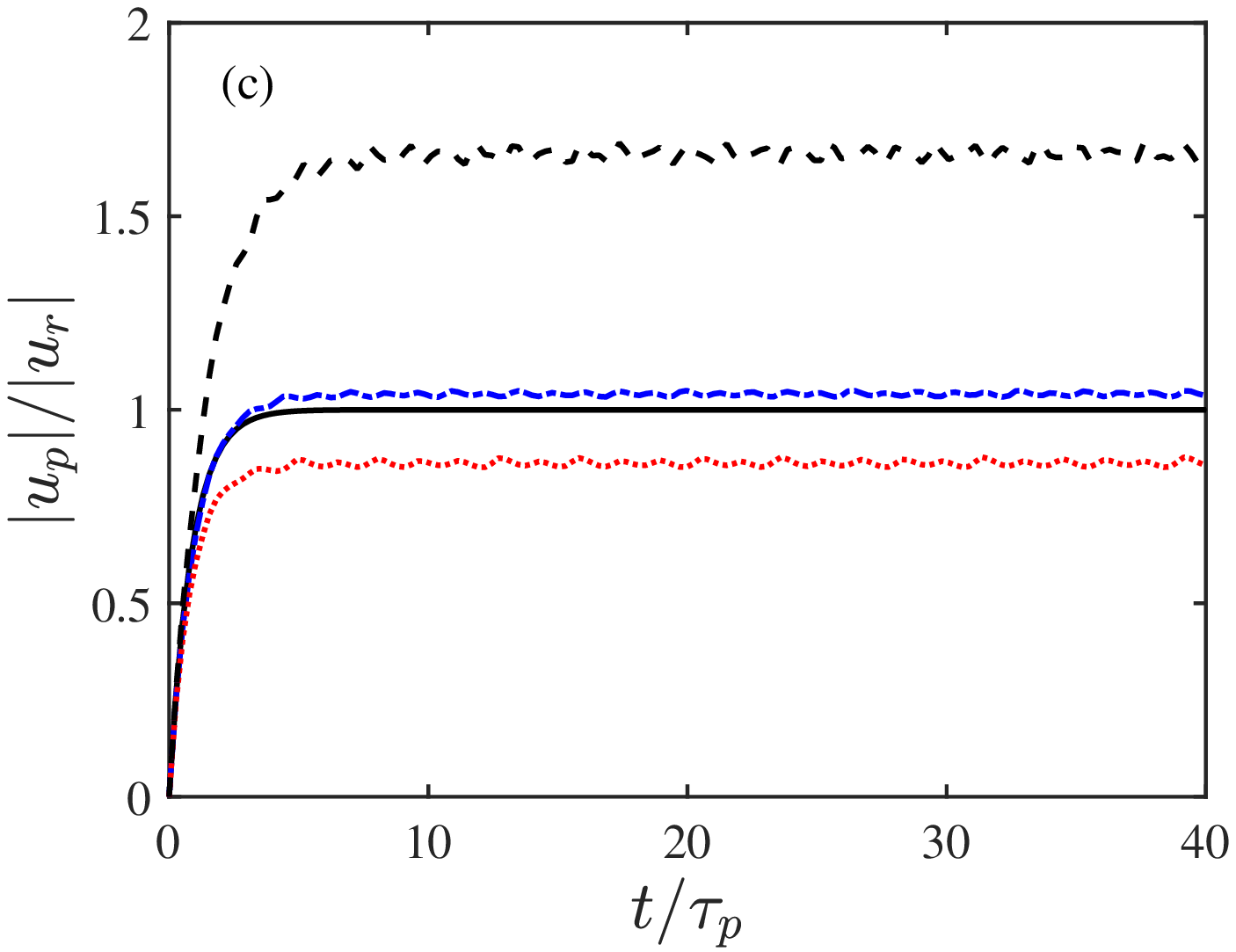}\hspace{0.04\textwidth}
    \includegraphics[scale=0.55]{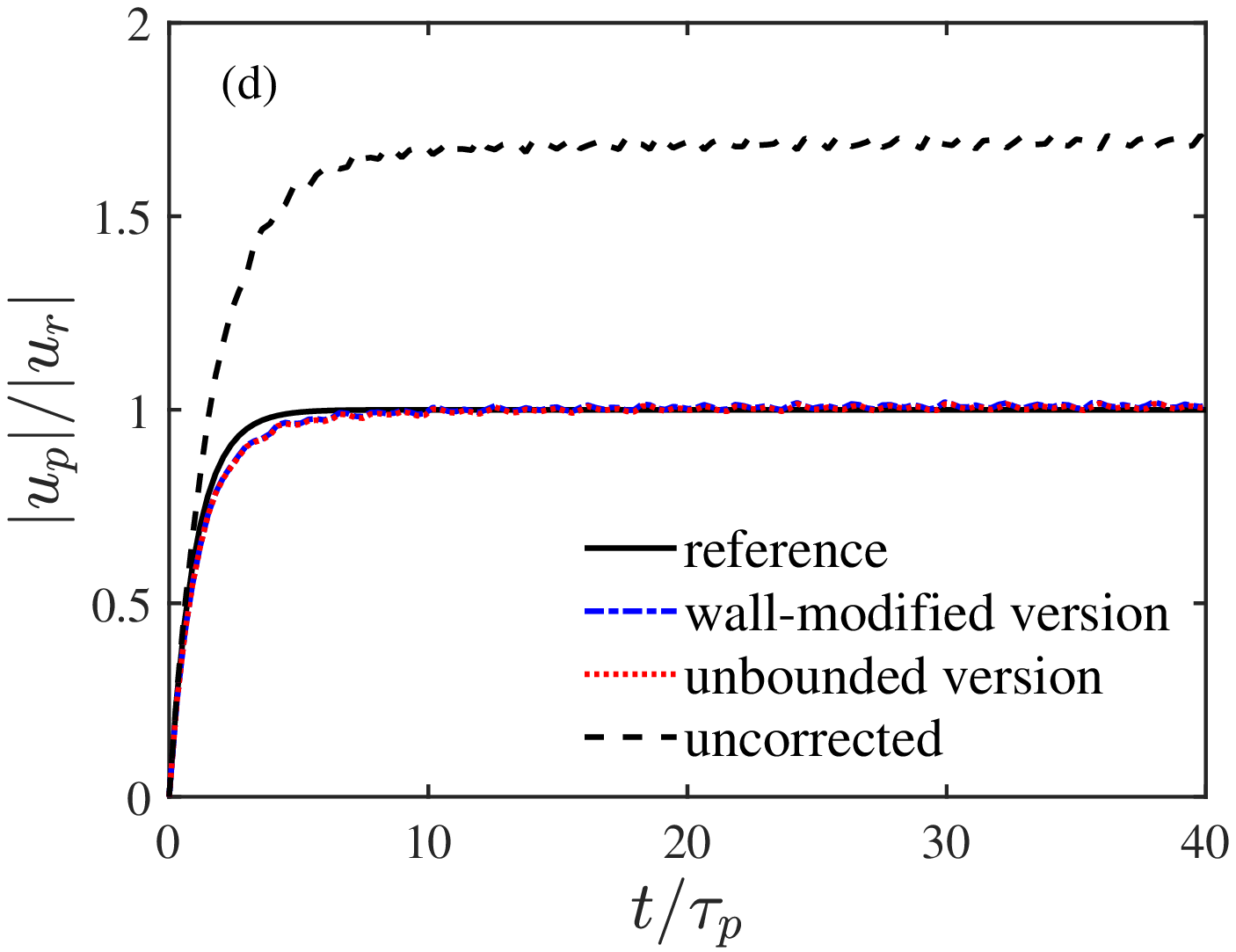}
    \caption{Shown are velocity of a particle settling under gravity parallel to a wall at different wall distances of (a): $\delta_p{=}0.05$, (b): $\delta_p{=}0.5$, (c): $\delta_p{=}1.5$ and (d): $\delta_p{=}\infty$. Results of the present scheme with wall-modified version (dash-dotted blue), unbounded version (dotted red) and uncorrected scheme (dashed black) are all compared against the reference velocity (solid black). These results are based on case A of Tab. \ref{tab:error_parallel}.}
    \label{fig:A_cases}
\end{figure}

When wall effects are accounted for in the correction scheme, the asymmetry pattern is captured which results in excellent predictions. For the cases considered, the errors reduce to one order of magnitude smaller values when the wall-modified correction scheme is applied. For example, in case A1, the total error of $91.42\%$ in particle settling velocity predicted by the unbounded correction scheme reduces to $6.03\%$ when wall effects are accounted for. Additionally, for particles travelling far away from the wall wherein the symmetric disturbance field is expected, the wall-modified and unbounded versions of the present correction scheme both yield nearly identical results. This shows the superiority of the former to the latter for general particle-laden flows. Figure \ref{fig:A_cases} illustrates the results of these two versions on the particle velocity of case A as a function of time. The erroneous results of the unbounded version for near wall motions is improved by including wall effects in the correction scheme. 

\begin{table}
\begin{center}
\def~{\hphantom{0}}
\begin{adjustbox}{width=\textwidth}
\begin{tabular} {lccccccc}
\hline
Case & $\delta_p$& $\Lambda^{(1)}$ & $\Lambda^{(2)}$ & $\Lambda^{(3)}$  &  \thead{uncorrected \\ $e^{\parallel}$ \quad \quad $e^{\perp}$ \quad \quad $e$} & \thead{corrected using \\ unbounded model\\ $e^{\parallel}$ \quad \quad $e^{\perp}$ \quad \quad $e$} & \thead{corrected using\\ wall-modified model\\ $e^{\parallel}$ \quad \quad $e^{\perp}$ \quad \quad $e$} \\ 
\hline
A1 & 0.05 & 1.0 & 1.0 & 1.0 & 125.82 \quad 0.17 \quad 125.82 & -86.82 \quad 21.42 \quad 91.42 & 5.37 \quad 2.23 \quad 6.03 \\ 
A2 & 0.5 & 1.0 & 1.0 & 1.0 & 59.16 \quad 0.095 \quad 59.16 & -35.08 \quad 1.37 \quad 35.12 & 4.86 \quad 0.57 \quad 4.91 \\
A3 & 1.0 & 1.0 & 1.0 & 1.0 & 103.12 \quad 0.073 \quad 103.12 & -19.67 \quad 1.16 \quad 19.72 & 4.29 \quad 0.76 \quad 4.38 \\
A4 & 1.5 & 1.0 & 1.0 & 1.0 & 66.12 \quad 0.073 \quad 66.12 & -13.81 \quad 0.66 \quad 13.84 & 4.06 \quad 0.46 \quad 4.10 \\
A5 & 2.0 & 1.0 & 1.0 & 1.0 & 102.96 \quad 0.06 \quad 102.96 & -10.02 \quad 0.87 \quad 10.08 & 2.08 \quad 0.72 \quad 2.24 \\
A6 & $\infty$ & 1.0 & 1.0 & 1.0 & 69.19 \quad 0.05 \quad 69.19 & 0.74 \quad 0.45 \quad 0.95 & 1.0 \quad 0.44 \quad 1.14 \\
&&&&&&&\\
B1 & 0.05 & 5.0 & 5.0 & 5.0 & 745.72 \quad 0.54 \quad 745.72 & -102.6 \quad 142.86 \quad 212.04 & -3.02 \quad 13.66 \quad 19.50 \\
B2 & 0.5 & 5.0 & 5.0 & 5.0 & 437.08 \quad 0.28 \quad 437.08 & -31.02 \quad 20.99 \quad 42.67 & 4.51 \quad 6.96 \quad 11.16 \\
B3 & 1.0 & 5.0 & 5.0 & 5.0 & 589.81 \quad 0.16 \quad 589.8 & -22.48 \quad 10.12 \quad 28.80 & -4.17 \quad 9.77 \quad 17.54 \\
B4 & 1.5 & 5.0 & 5.0 & 5.0 & 390.42 \quad 0.22 \quad 390.42 & -10.43 \quad 6.02 \quad 15.35 & 4.78 \quad 5.58 \quad 10.01 \\
B5 & 2.0 & 5.0 & 5.0 & 5.0 & 554.19 \quad 0.15 \quad 554.19 & -9.99 \quad 8.89 \quad 20.14 & -0.59 \quad 9.08 \quad 16.06 \\
B6 & $\infty$ & 5.0 & 5.0 & 5.0 & 353.79 \quad 0.2 \quad 353.79 & 0.35 \quad 5.01 \quad 9.16 & 0.83 \quad 4.85 \quad 8.97 \\
&&&&&&&\\
C1 & 0.05 & 0.1 & 1.0 & 0.2 &  7.91 \quad 0.17 \quad 7.91 & -33.79 \quad 2.12 \quad 33.87 & 0.58 \quad 0.33 \quad 0.67 \\
C2 & 0.5 & 0.1 & 1.0 & 0.2 & 5.59 \quad 0.25 \quad 5.59 & -18.12 \quad 1.14 \quad 18.16 & 1.03 \quad 0.32 \quad 1.08 \\
C3 & 1.0 & 0.1 & 1.0 & 0.2 & 9.82 \quad 0.46 \quad 9.83 & -11.47 \quad 0.78 \quad 11.51 &  1.29 \quad 0.53 \quad 1.40 \\
C4 & 1.5 & 0.1 & 1.0 & 0.2 & 8.93 \quad 0.58 \quad 8.95 & -11.27 \quad 0.62 \quad 11.29 & 0.71 \quad 0.59 \quad 0.93 \\
C5 & 2.0 & 0.1 & 1.0 & 0.2 & 11.98 \quad 0.77 \quad 12.01 & -7.55 \quad 0.42 \quad 7.57 & 0.55 \quad 0.65 \quad 0.86 \\
C6 & $\infty$ & 0.1 & 1.0 & 0.2 & 14.88 \quad 1.09 \quad 14.92 & -2.16 \quad  0.25 \quad 2.19 & -1.89 \quad 0.25 \quad 1.92 \\
&&&&&&&\\
D1 & 0.05 & 0.5 & 5.0 & 1.0 & 106.87 \quad 7.99\quad 107.17 & -69.94 \quad 3.33 \quad 70.02 & -3.04 \quad 7.25 \quad 8.40 \\
D2 & 0.5 & 0.5 & 5.0 & 1.0 & 82.44 \quad 7.69 \quad 82.80 & -44.12 \quad 1.03 \quad 44.15 & -12.55 \quad 4.7 \quad 13.45 \\
D3 & 1.0 & 0.5 & 5.0 & 1.0 & 95.70 \quad 8.00 \quad 96.04 & -16.49 \quad 1.93 \quad 16.64 & -4.21 \quad 2.72 \quad 5.05 \\
D4 & 1.5 & 0.5 & 5.0 & 1.0 & 82.49 \quad 7.24 \quad 82.81 & -20.79 \quad 1.17 \quad 20.84 & -9.87 \quad 2.14 \quad 10.23 \\
D5 & 2.0 & 0.5 & 5.0 & 1.0 & 94.04 \quad 7.12 \quad 94.31 & -9.59 \quad 0.84 \quad 9.64 & -0.78 \quad 1.58 \quad 3.13 \\
D6 & $\infty$ & 0.5 & 5.0 & 1.0 &  79.99 \quad 5.95 \quad 80.21 & -10.84 \quad 1.01 \quad 10.92 & -9.53 \quad 0.99 \quad 9.63   \\
&&&&&&&\\
E1 & 0.05 & 0.3 & 6.0 & 0.6 & 42.99 \quad 2.14 \quad 43.05 & -43.24 \quad 2.93 \quad 43.34  & -0.69 \quad 1.54 \quad 1.71\\
E2 & 0.5 & 0.3 & 6.0 & 0.6 & 50.17 \quad 3.78 \quad 50.31 & -21.67 \quad 0.9 \quad 21.69 & -3.15 \quad 1.34 \quad 3.45 \\
E3 & 1.0 & 0.3 & 6.0 & 0.6 & 49.20 \quad 4.28 \quad 49.39 & -16.29 \quad 0.14 \quad 16.29 & -4.33 \quad 1.40 \quad 4.58 \\
E4 & 1.5 & 0.3 & 6.0 & 0.6 & 48.27 \quad 4.34 \quad 48.47 & -12.23 \quad 0.91 \quad 12.28  & -4.43 \quad 1.77 \quad 4.88 \\
E5 & 2.0 & 0.3 & 6.0 & 0.6 &  56.16 \quad 5.08 \quad 56.39 & -9.27 \quad 0.65 \quad 9.29 & -3.47 \quad 1.2 \quad 3.68 \\
E6 & $\infty$ & 0.3 & 6.0 & 0.6 & 53.61 \quad 4.05 \quad 53.77 & -5.47 \quad 0.23 \quad 5.48 & -4.41 \quad 0.31 \quad 4.42 \\
&&&&&&&\\
F1 & 0.05 & 0.6 & 12.0 & 1.2 & 113.56 \quad 8.25 \quad 113.86 & -50.01 \quad 1.61 \quad 50.03 & -4.82 \quad 4.09 \quad 6.52\\
F2 & 0.5 & 0.6 & 12.0 & 1.2 & 121.47 \quad 11.00 \quad 121.97 & -19.90 \quad 1.18 \quad 19.94 & -3.06 \quad 3.01 \quad 4.44 \\
F3 & 1.0 & 0.6 & 12.0 & 1.2 & 113.10 \quad 9.81 \quad 113.53 & -12.64 \quad 1.51 \quad 12.74 & -3.03 \quad 2.47 \quad 4.2 \\
F4 & 1.5 & 0.6 & 12.0 & 1.2 & 108.30 \quad 8.8 \quad 108.66 & -10.05 \quad 1.25 \quad 10.13 & -3.48 \quad 1.86 \quad 4.27\\
F5 & 2.0 & 0.6 & 12.0 & 1.2 & 105.95 \quad 8.24 \quad 106.27 & -8.19 \quad 1.01 \quad 8.26 & -3.28 \quad 1.44 \quad 3.82 \\
F6 & $\infty$ & 0.6 & 12.0 & 1.2 & 100.45 \quad 7.42 \quad 100.72 & -6.05 \quad 0.72 \quad 6.10  & -4.12 \quad 0.90 \quad 4.27 \\
\hline
\end{tabular}
\end{adjustbox}
\caption{Tabulated are the percentage errors in the simulated velocity of a single particle settling parallel to a wall under gravity and at different normalized wall gaps. Different sets of computations including various types of grid aspect ratio as well as particle diameter to the grid size, $\Lambda^{(i)}$ are studied. For each set, different wall distances of $\delta_p$, is examined to study the error in the settling velocity, $e^{||}$, drifting velocity, $e^{\perp}$, and the overall error, $e$. Flow parameters are kept constant in all cases with Stokes number of $St{=}10$ and unbounded particle Reynolds number of $Re^{Stk}_p{=}0.1$. The results of the present wall-modified correction scheme is compared with its unbounded counterpart as well as the classical uncorrected point-particle approach.}
\label{tab:error_parallel}
\end{center}
\end{table}

The results presented in the previous part were obtained for $Re_p{<}0.1$, while in the wall-bounded particle-laden flows, typically a wider range of $Re_p$ exists. In this part, the present model is tested for a range of $Re_p$ up to 10 by performing similar computations to the previous part. Table \ref{tab:error_Re_St} lists the studied cases for this part that are similar to case E1 of Tab. \ref{tab:error_parallel}, yet with different Stokes and particle Reynolds numbers. Unlike the previous part, the reported particle Reynolds number here is based on its actual velocity and defined by $Re_p{=}Re^{Stk}_p/f^{||}(\delta_p,Re_p)$ which varies from 0.044 to 10. For all cases, settling is performed at $\delta_p{=}0.05$ for which the deviation between unbounded and wall-modified correction schemes of the previous part was found to be significant. For studied cases here, the whole terms in Eq. \ref{eq:f_parallel} hold and we use first order forward Euler finite difference scheme to solve Eq. \ref{eq:up_parallel} and obtain its reference velocity as a function of time. 

As shown in Tab. \ref{tab:error_Re_St}, the error in uncorrected scheme is reduced as $Re_p$ increases which is in line with the observations of the preceding works \citep{horwitz2018,balachandar2019}. This is conceptually justified due to the fact that, unlike particles with small $Re_p$, higher Reynolds number particles move faster and stay less in their own disturbance field created in the previous time step. Although this diminishes the need for the correction, the error of approximately $30\%$ that pertains to the case with $Re_p{=}10$ (the largest studied $Re_p$), is still considerable. As listed in Tab. \ref{tab:error_Re_St}, the present wall-modified correction scheme reduces the errors by approximately one order of magnitude for cases with $Re_p{<}10$ and results in better predictions compared to the unbounded version wherein the wall effects are ignored. 

It should be emphasized that the present model is constructed based on the small $Re_p$ assumption. Although the finite $Re_p$ effects are partially accounted for through the factor $C_r$ (Eq. \ref{eq:cr}), a more elaborate formulation is required to improve the accuracy of the model for $Re_p{>}10$. For such cases, the assumption of symmetric Stokes solution is not valid anymore and the linear superposition of the perturbations caused by neighbour cells used in the  derivation of $K_p$ may be broken, that are left for future investigations. Similar observations were achieved by \cite{horwitz2018} wherein they showed that their unbounded correction scheme that was developed based on small $Re_p$, is still reliable for cases with $Re_p$ of 10 with errors in settling velocity of $10\%$. Concerning the Stokes number effects, our results show insignificant changes to the prediction of the present model for the studied range of this parameter ($3{<}St{<}30$).

\begin{table}
\begin{center}
\def~{\hphantom{0}}
\begin{adjustbox}{width=0.9\textwidth}
\begin{tabular} {lccccccccccc}
\hline
Case & $Re_p$ & $St$ & \thead{uncorrected \\ $e^{\parallel}$ \quad \quad $e^{\perp}$ \quad \quad $e$} & \thead{corrected using \\ unbounded version\\ $e^{\parallel}$ \quad \quad $e^{\perp}$ \quad \quad $e$} & \thead{corrected using\\ wall-modified version\\ $e^{\parallel}$ \quad \quad $e^{\perp}$ \quad \quad $e$} \\% & \thead{corrected using\\ new wall-modified model\\ $e^{\parallel}$ \quad \quad $e^{\perp}$ \quad \quad $e$}\\ 
\hline

R1 & 0.044 & 3.0 &  32.94 \quad 1.64 \quad 32.99 & -41.59 \quad 2.77 \quad 41.68 & -3.10 \quad 0.77 \quad 3.20 \\ %& -3.38 \quad 0.77 \quad 3.48\\
R2 & 0.044 & 10.0 & 40.80 \quad 2.02 \quad 40.85 & -41.58 \quad 2.84 \quad 41.68 & -2.74 \quad 0.86 \quad 2.88 \\ %& -3.16 \quad 0.84 \quad 3.27\\
R3 & 0.044 & 30.0 & 61.31 \quad 3.09 \quad 61.38 & -47.91 \quad 3.63 \quad 48.05 & -6.75 \quad 1.23 \quad 6.87 \\ %& -7.75 \quad 1.19 \quad 7.86\\
&&&&&& \\

R4 & 0.5 & 3.0 & 50.95 \quad 2.60 \quad 51.02 & -48.35 \quad 3.99 \quad 48.52 & -4.11 \quad 1.43 \quad 4.66 \\% & -10.61 \quad 1.28 \quad 10.69\\
R5 & 0.5 & 10.0 & 53.11 \quad 2.68 \quad 53.18 & -49.24 \quad 3.74 \quad 49.41 & -1.86 \quad 1.31 \quad 2.78 \\% & -6.54\quad 1.19 \quad 6.70 \\
R6 & 0.5 & 30.0 & 52.43 \quad 2.62 \quad 52.50 & -47.14 \quad 3.07 \quad 47.26 & -1.46 \quad 1.24 \quad 2.11 \\ %& -5.36 \quad 1.13 \quad 5.49 \\
&&&&&& \\

R7 & 5.0 & 3.0 & 39.62 \quad 1.77 \quad 39.66 & -26.45 \quad 2.34\quad 26.58 & 2.76 \quad 1.11 \quad 3.18 \\% & -9.49 \quad 0.91 \quad 9.55\\
R8 & 5.0 & 10.0 & 39.77 \quad 1.76 \quad 39.81 & -26.18 \quad 2.40 \quad 26.31 & 2.96 \quad 1.10 \quad 3.23 \\ %& -9.54 \quad 0.91 \quad 9.60\\
R9 & 5.0 & 30.0 & 39.90 \quad 1.76 \quad 39.94 & -26.16 \quad 2.37 \quad 26.27 & 3.04 \quad 1.10 \quad 3.24 \\% & -9.36 \quad 0.91 \quad 9.40\\
&&&&&& \\

R10 & 10.0 & 3.0 & 33.56 \quad 1.26 \quad 33.59 & -17.60 \quad 2.23 \quad 17.77 & 5.25 \quad 0.76 \quad 5.31 \\% & -8.50 \quad 0.55 \quad 8.55 \\
R11 & 10.0 & 10.0 & 34.06 \quad 1.27 \quad 34.09 & -17.45 \quad 2.26 \quad 17.60 & 5.45 \quad 0.76 \quad 5.50 \\%& -8.27 \quad 0.55 \quad 8.30 \\
R12 & 10.0 & 30.0 & 33.92 \quad 1.26 \quad 33.94 & -17.39 \quad 2.27 \quad 17.54 & 5.40 \quad 0.74 \quad 5.45 \\%& -8.22 \quad 0.53 \quad 8.23 \\
&&&&&& \\

\hline
\end{tabular}
\end{adjustbox}
\caption{The effects of particle Reynolds number, $Re_p$, and particle Stokes number, $St$, on the velocity of a single particle settling parallel and close to a wall at $\delta_p{=}0.05$ are shown. The anisotropic grid resolution of case E from Tab. \ref{tab:error_parallel} with $\mathbf{\Lambda}{=}[0.3,6.0,0.6]$ is employed for all cases. The wall-modified and unbounded versions of the present correction scheme are compared together and against the uncorrected PP approach in terms of the error in settling velocity, $e^{||}$, drifting velocity, $e^{\perp}$ and total velocity, $e$.}
\label{tab:error_Re_St}
\end{center}
\end{table}

%%%%%%%%%%%%%%%%%%%%%%%%%%%%%%%%%%%%%%%%%%%%%%%%%%%%%
\subsection{Free falling particle normal to the wall}
%%%%%%%%%%%%%%%%%%%%%%%%%%%%%%%%%%%%%%%%%%%%%%%%%%%%
This section verifies the present model for capturing the disturbance field in the wall-normal motion of particles, as commonly encountered in wall-bounded particle-laden flows. The free falling motion of a particle normal to the wall is considered as a test case for this part. In such scenario, a particle falls under gravity and its drag coefficient increases as it approaches to the wall, owing to the wall lubrication effects. \cite{gondret1999} observed that depending on the particle Stokes number, it could either sit on the wall if $St{<}20$ or hit the wall and re-bound if $St{>}20$. To eliminate the particle-wall collision and isolate the particle-fluid interaction only, we perform the first situation wherein the particle is supposed to retard and sit on the wall. Accordingly, the Stokes number of $St{=}10$ is chosen for all the studied cases of this part.

\cite{brenner1961} derived an exact solution for the wall adjustment to the drag coefficient of a particle in normal motion to the wall which has small Reynolds number of $Re_p{<}0.1$. In their work, a corresponding asymptotic solution was also obtained that matches their exact solution for the normalized wall gaps of $\delta_p{>}1.38$. For $\delta_p{<}1.38$, \cite{cox1967} achieved an asymptotic solution that combined with the one obtained by \cite{brenner1961} are used in this work for the wall adjustment drag coefficient of a particle in normal motion toward the wall. This adjustment is expressed as 

\begin{equation}
f^{\perp}(\delta_p) = 
\begin{cases}
 1 +\left( \frac{0.562}{1+2\delta_p}\right), \quad \delta_p>1.38 \quad \text{\citep{brenner1961}} \\ 
 \frac{1}{2\delta_p} \left( 1+ 0.4\delta_p\log\left(\frac{1}{2\delta_p}\right) + 1.94\delta_p \right), \quad \delta_p<1.38 \quad \text{\citep{cox1967}}
\end{cases}
\label{eq:brener_asymptotic}
\end{equation}

\begin{figure}
    \centering
    \includegraphics[scale=0.6]{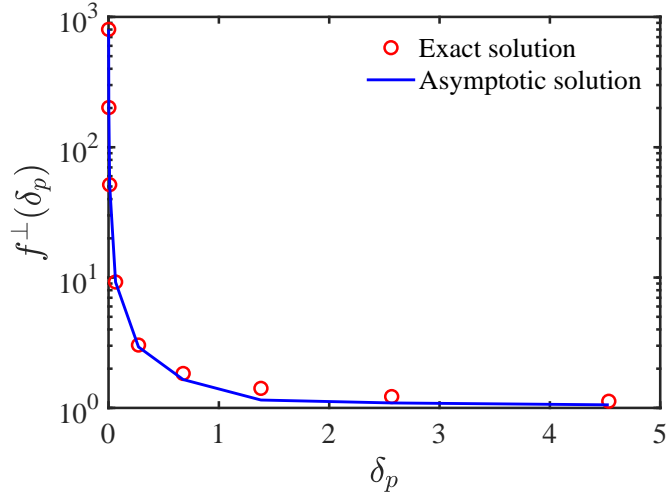}
    \caption{Shown is the wall adjustment to the drag coefficient of a particle in wall-normal motion. Exact solution of \protect\cite{brenner1961} is shown along with the asymptotic solution provided by \protect\cite{brenner1961} and \protect\cite{cox1967}, given in Eq. \ref{eq:brener_asymptotic}.}
    \label{fig:brener}
\end{figure}

Figure \ref{fig:brener} compares these asymptotic solutions to the exact solution of \cite{brenner1961}. Based on the adjustment factor provided by Eq. \ref{eq:brener_asymptotic}, the particle equation of motion (Eq. \ref{eq:dup_dt}) is solved for the reference velocity using a first order forward Euler finite difference scheme. For all the studied cases, the particle is initially stationary and located at the normalized wall gap of $\delta_p{=}7$ and falls under gravity. Similar to the preceding section, results of the wall-modified and unbounded versions of the present correction scheme are compared with those of the uncorrected approach. Studied cases are listed in Tab. \ref{tab:error_normal} that are carried out using both isotropic and anisotropic grids. A range of particle Reynolds number of $0.04{<}Re_p{<}10$ and Stokes number of $3{<}St{<}30$ are used for each grid resolution. For each case, the total time that particle requires to reach the normalized wall gap of $\delta_p{=}0.5$ is computed and compared against the corresponding reference value, $t_{ref}$. The deviation of each scheme from the reference is quantified based on the following metric

\begin{equation}
    e = \frac{t- t_{ref}}{t_{ref}}
    \label{error_free_fall}
\end{equation}

As Tab. \ref{tab:error_normal} shows, without correcting the PP approach, the considerable and negative errors for each case imply that particle sees a smaller drag force due to the disturbance created in the background flow, accelerates faster and reaches the wall-gap of interest quicker. However, when the PP is corrected using the present wall-modified correction scheme, it reduces the errors and results in better prediction for the particle trajectory and velocity. Although the errors obtained based on the unbounded version of the correction scheme are still better than the uncorrected approach, the superiority of the wall-modified version on other schemes is observed in this case as well. 

\begin{table}
\begin{center}
\def~{\hphantom{0}}
\begin{adjustbox}{width=0.9\textwidth}
\begin{tabular} {lcccccccc}
\hline
Case & $Re^{Stk}_p$ & $St$ &  $\Lambda^{(1)}$ & $\Lambda^{(2)}$ & $\Lambda^{(3)}$  &  \thead{uncorrected \\ $e$} & \thead{corrected using \\ unbounded version\\ $e$} & \thead{corrected using \\ wall-modified version \\ $e$} \\ 
\hline
N1 & 0.1 & 3.0 & 1.0 & 1.0 & 1.0 & -30.55 & 32.30 & -6.05\\
N2 & 0.1 & 10.0 & 1.0 & 1.0 & 1.0 & -24.37 & 23.86 & -4.77\\
N3 & 0.1 & 30.0 & 1.0 & 1.0 & 1.0 & -15.09 & 10.13 & -2.79 \\
N4 & 5.0 & 10.0 & 1.0 & 1.0 & 1.0 & -2.30 & 0.97 & -0.07\\
N5 & 10.0 & 10.0 & 1.0 & 1.0 & 1.0 & -1.37 & 0.62 & 0.02\\
&&&&&&&&\\

N6 & 0.1 & 3.0 & 0.3 & 6.0 & 0.6 & -9.62 & 4.95 & -2.88\\
N7 & 0.1 & 10.0 & 0.3 & 6.0 & 0.6 & -9.61  & 5.04 & -2.75\\
N8 & 0.1 & 30.0 & 0.3 & 6.0 & 0.6 & -9.57 & 4.78 & -2.87\\
N9 & 5 & 10.0 & 0.3 & 6.0 & 0.6 & -1.80 & 0.72 & -0.24\\
N10 & 10 & 10.0 & 0.3 & 6.0 & 0.6 & -0.73 & 0.52 & 0.1\\
%N11 & 0.1 & 10.0 & 0.5 & 5.0 & 1.0 & -20.32 & 7.96 & -5.08 \\
%N12 & 0.1 & 10.0 & 0.6 & 12.0 & 1.2 & -24.62 & -0.96 & -9.62\\
\hline
\end{tabular}
\end{adjustbox}
\caption{Errors calculated in the prediction of particle's wall-normal motion. Two sets of grid aspect ratio with various particle Reynolds numbers and Stokes numbers are performed. For each case, the error in the time that particle requires to reach the normalized wall gap of $\delta_p{=}0.5$ is computed based on the wall-modified and unbounded versions of the present correction scheme in comparison with that of the uncorrected scheme.}
\label{tab:error_normal}
\end{center}
\end{table}

\begin{figure}
    \centering
    \includegraphics[scale=0.545]{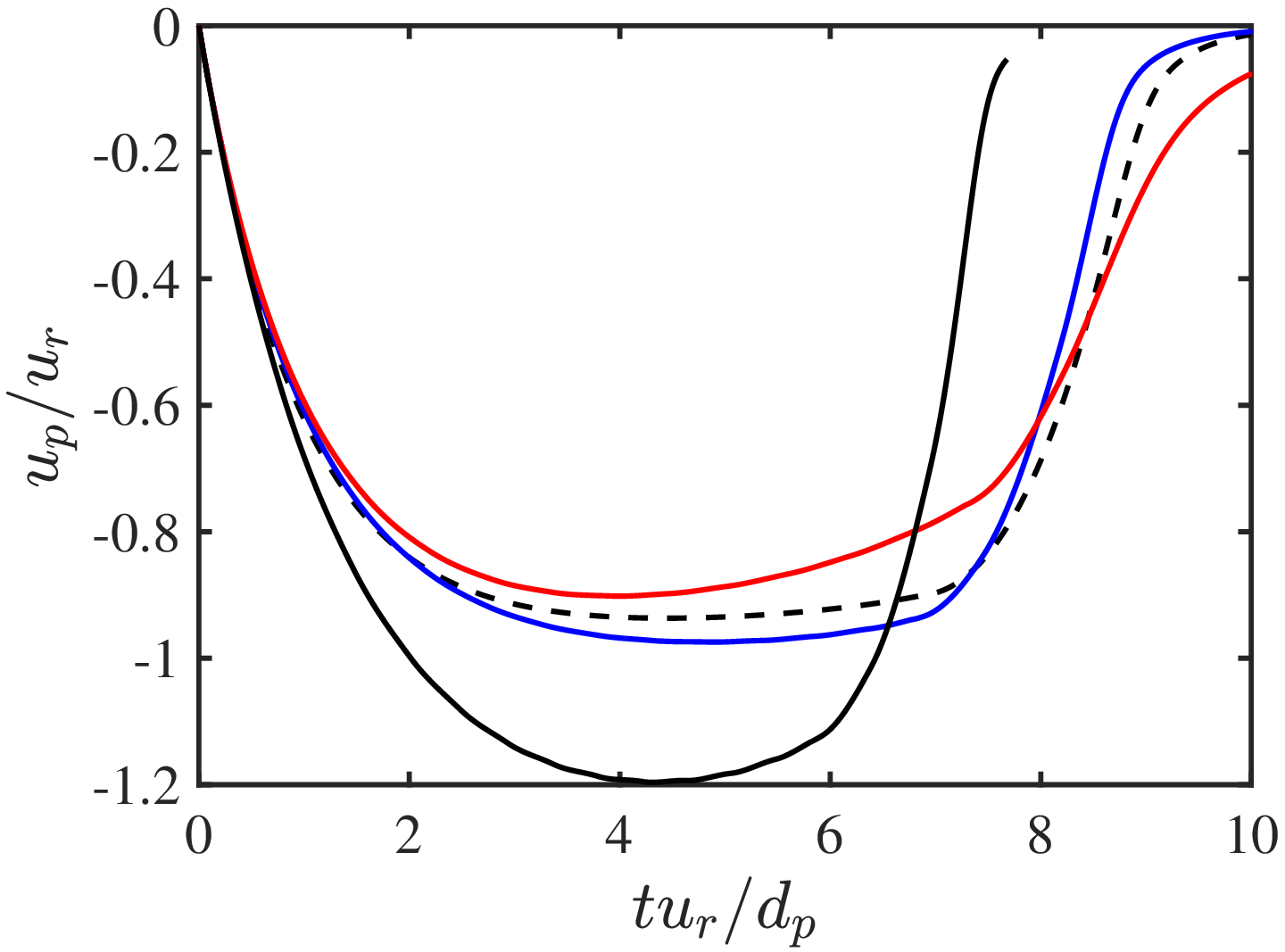}\hspace{0.035\textwidth}
    \includegraphics[scale=0.545]{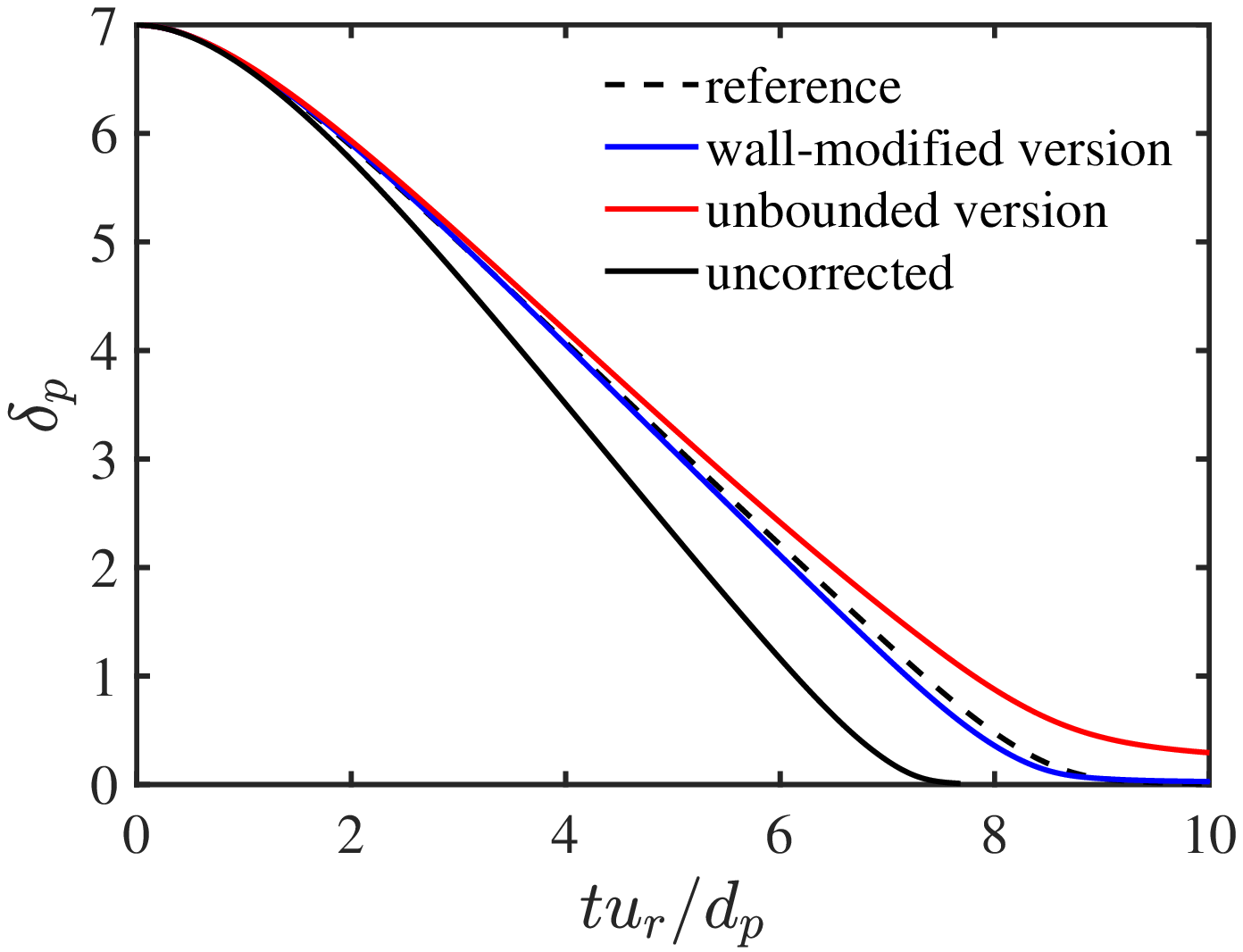}
    \caption{Shown are the normalized velocity (left) and normalized wall-gap (right) of a particle settling normal to a wall predicted by the wall-modified and unbounded versions of the present correction scheme. Results are compared with the uncorrected scheme as well as the reference. Results here pertain to the case N2 from Tab. \ref{tab:error_normal}. }
    \label{fig:isotropic_normal}
\end{figure}

Figure \ref{fig:isotropic_normal} shows the prediction of different schemes on the particle velocity and trajectory of case N2 from Tab. \ref{tab:error_normal}. The reference velocity used for normalization is based on Eq. \ref{eq:u_set_stokes} that pertains to the Stokes settling velocity of a particle in an unbounded domain. As illustrated, the wall-modified version of the present model captures quite well the accurate trajectory and velocity of the particle whereas the unbounded scheme hinders the particle settling due to the overprediction in the disturbance field. Results in this part along with the observation of the previous parts underscore the need for accounting for the wall effects in capturing the disturbance field and having a general correction scheme that could be applied to all types of particle-laden flows in the presence and absence of no-slip boundaries.   

%%%%%%%%%%%%%%%%%%%%%
\section{Conclusion}
%%%%%%%%%%%%%%%%%%%%%
\label{sec:conclusion}
Modeling two-way coupled Euler-Lagrange (EL) particle-laden flows using point-particle (PP) approach can result in erroneous predictions due to an issue that arises in the calculation of fluid forces acting on the particles. The available closures for force calculations are all based on the undisturbed fluid velocity, which by definition is the fluid velocity not influenced by the presence of particles. In the two-way coupled computations, however, the particle reaction force disturbs the fluid velocity around the particle and using such a disturbed velocity for force calculations in the next time step, yields inaccurate inter-phase interactions and wrong predictions. More importantly, depending on whether the particle is travelling near a no-slip boundary or in an unbounded domain, its disturbance in the background flow can be different in terms of shape and strength, and can also be asymmetric.

In this paper, we presented a general correction scheme for EL-PP approaches to recover the undisturbed fluid velocity from the available disturbed field in the presence and absence of the no-slip walls. %The model is an extension and generalized version of the correction scheme by \cite{esmaily2018} that was originally introduced for the unbounded regimes and in the absence of the no-slip walls. 
In the present velocity correction approach, the disturbance created by a particle in a computational cell that carries the particle is obtained by finding the response of the cell (its velocity) to the particle force. Analogous to the motion of a solid object, the disturbance velocity of the computational cell is obtained by treating the computational cell as a solid object that is subjected to the particle force and dragged through the adjacent computational cells \citep{esmaily2018}. Knowing these two forces, the disturbance velocity of the cell is solved using a Maxey-Riley equation of motion for the computational cell. The model is general and can be used for (i) unbounded and wall-bounded regimes, (ii) isotropic and anisotropic grid resolutions, (iii) particles bigger than the grid size, (iv) arbitrary interpolation and distribution functions, and (v) flows with finite particle Reynolds number. 

An empirical expression was obtained for the drag coefficient of the computational cell ($K^{(i)}_c$) that is applicable for a wide range of grid aspect ratios, typically encountered in the particle-laden turbulent channel flows. The new expression, obtained based on the employed collocated grid arrangement, is a function of the grid size. Just as a slowly moving solid particle in a quiescent fluid influences the near field through Stokes solution, the particle force at a computational cell perturbs the surrounding cells. It was shown that for the employed collocated grid arrangement, Stokes solution normalized by the characteristics length scale of $0.25d_c$ results in accurate predictions for the disturbance field created in the surrounding cells in comparison with the numerical measurements.

%The disturbance normalized by the characteristics length scale of $0.25d_c$ results in much more accurate numerical measurements.

%and is applicable for a wide range of grid aspect ratios
%and was shown to be more accurate than the one used in \cite{esmaily2018}, for high aspect ratios. \cite{esmaily2018} showed that the computational cell subjected to the particle force perturbs its surrounding cells in the same way that a slow moving solid sphere influences its adjacent quiescent fluid through the Stokes solution. Using staggered grid arrangement, they found that this solution normalized with the characteristic length of $0.28d_c$ predicts best their numerical measurements. For the collocated grid arrangement employed here, however, we observed that the choice of $0.25d_c$ results in better predictions based on our numerical measurements. 

Wall effects in the model were taken into account through two different factors; (i) $\Psi^{(i)}_k$ and (ii) $\Phi^{(i)}_{kj}$. The first pertains to the wall modification to the drag coefficient of the computational cell near a no-slip boundary, analogous to the near wall motion of a solid object. Two components for this parameter were obtained for the disturbances created in parallel and wall-normal directions. For isotropic grid, it was shown that the wall adjustment to the drag coefficient of a solid sphere moving near a no-slip wall, empirically derived by \cite{zeng2009}, can be an excellent choice for $\Psi^{(i)}_k$. However, for anisotropic grids owing to their large aspect ratios, this expression does not hold, and a new fitted expression was obtained that covers a wide range of grid sizes and aspect ratios. The second parameter, $\Phi^{(i)}_{kj}$, was introduced to capture the wall effect on the Stokes solution of the computational cell. It was shown that perturbation created at neighbouring cells by a computational cell that is exposed to the particle force differs in shape and strength as the cell becomes closer to a no-slip wall. It was argued that one could directly use the wall-modified Stokes solution instead of its unbounded counterpart, however, due to the complexity and expense embedded in the implementation and solution of the wall-modified version, Stokeslet solution was suggested as the second wall adjustment factor. In that regard, we kept the Stokes solution in the formulation, while its wall effect was accounted for by multiplying this solution by the ratio of the wall-bounded to the unbounded Stokeslet solutions, defined as $\Psi^{(i)}_k$. Our results showed that the choice of this ratio yields in good predictions with small errors. 

An unbounded version of the present model can be obtained by letting $\Psi^{(i)}_k{=}\Phi^{(i)}_{kj}{=}1$ in the formulation, that can be used in particle-laden flows without no-slip boundary conditions. To verify the collocated adjustments made in the formulation, the unbounded version of the scheme was first tested for settling of a particle in an unbounded domain and results were compared with those reported in \cite{esmaily2018}. For the different studied flow and grid parameters, it was shown that the model using the collocated grid arrangement accurately captures the settling velocity of the particle with a few percent errors. 

To assess the model for wall-bounded applications, settling of a particle parallel to a no-slip wall was performed at various wall-normal distances. Consistent with the observation of \cite{esmaily2018}, the error in the uncorrected particle velocity was observed to be a function of particle's diameter to the grid size, $(d_p/d_c)$. Correcting the PP approach with the current model, however, captured the disturbance field at all wall distances and significantly reduced the errors in the predicted particle velocity by accurately recovering the undisturbed field. Furthermore, it was observed that ignoring the wall effects in the formulation for wall-bounded flows, i.e., assuming $\Psi^{(i)}_k{=}\Phi^{(i)}_{kj}{=}1$, results in large errors that are in the same order of magnitude of the uncorrected scheme, particularly in the near wall motions. As particle gets away from the wall, however, the effects of wall diminish and the formulation approaches the unbounded version. 

Tests performed for a range of $0{<}Re_p{<}10$ revealed the fact that the error in the uncorrected settling velocity decreases as $Re_p$ increases, consistent with the observation of \cite{balachandar2019}. This is justified due to the fact that particles with large $Re_p$ do not stay in their own disturbance, created in the previous time step, and this alleviates the need for the correction. Nevertheless, the relatively small errors associated with large $Re_p$ cases was still lowered using the present correction scheme.  

The last test cases were carried out on the free falling motion of a particle in the wall-normal direction. It was shown that the particle's velocity in the uncorrected scheme is erroneously overpredicted which makes the particle hit the wall earlier than it would in reality. When the PP approach is corrected with the present model, however, it recovers the undisturbed velocity at any wall distance and captures the particle's velocity and trajectory more accurately. Tests performed for this part with different grid configurations and flow parameters showed the superiority of the present model to the uncorrected and unbounded correction schemes. 

The present correction scheme is general, cost-efficient and accurate that can be easily implemented in EL-PP packages to study a wide range of particulate flows with and without the no-slip boundaries. We conjecture that this scheme could help improve the investigations and the state-of-the-art of the wall-bounded particle-laden flows wherein the lack of accuracy of the standard uncorrected PP approaches has been widely observed. For such flows, the proposed correction scheme can significantly improve the predictive capability of point-particle method approaching those of the particle-resolved methods at significantly lower computational cost.

%%%%%%%%%%%%%%%%%%%%%%%%%
\section{Acknowledgements}
%%%%%%%%%%%%%%%%%%%%%%%%%
Financial support was provided under the NASA Contract Number NNX16AB07A monitored by program manager Dr. Jeff Moder, NASA Glenn Research Center as well as the National Science Foundation (NSF) under Grant No. 1133363. In addition, the authors acknowledge the San Diego Supercomputer Center (SDSC) at University of California San Diego for providing HPC resources that have contributed to the results reported here.

%%%%%%%%%%%%%%%%%%
\section*{Appendix A. A simplified equation for $\Psi^{(i)}_k$ on isotropic grids}
\begin{figure}
    \centering
    \includegraphics[scale=0.55]{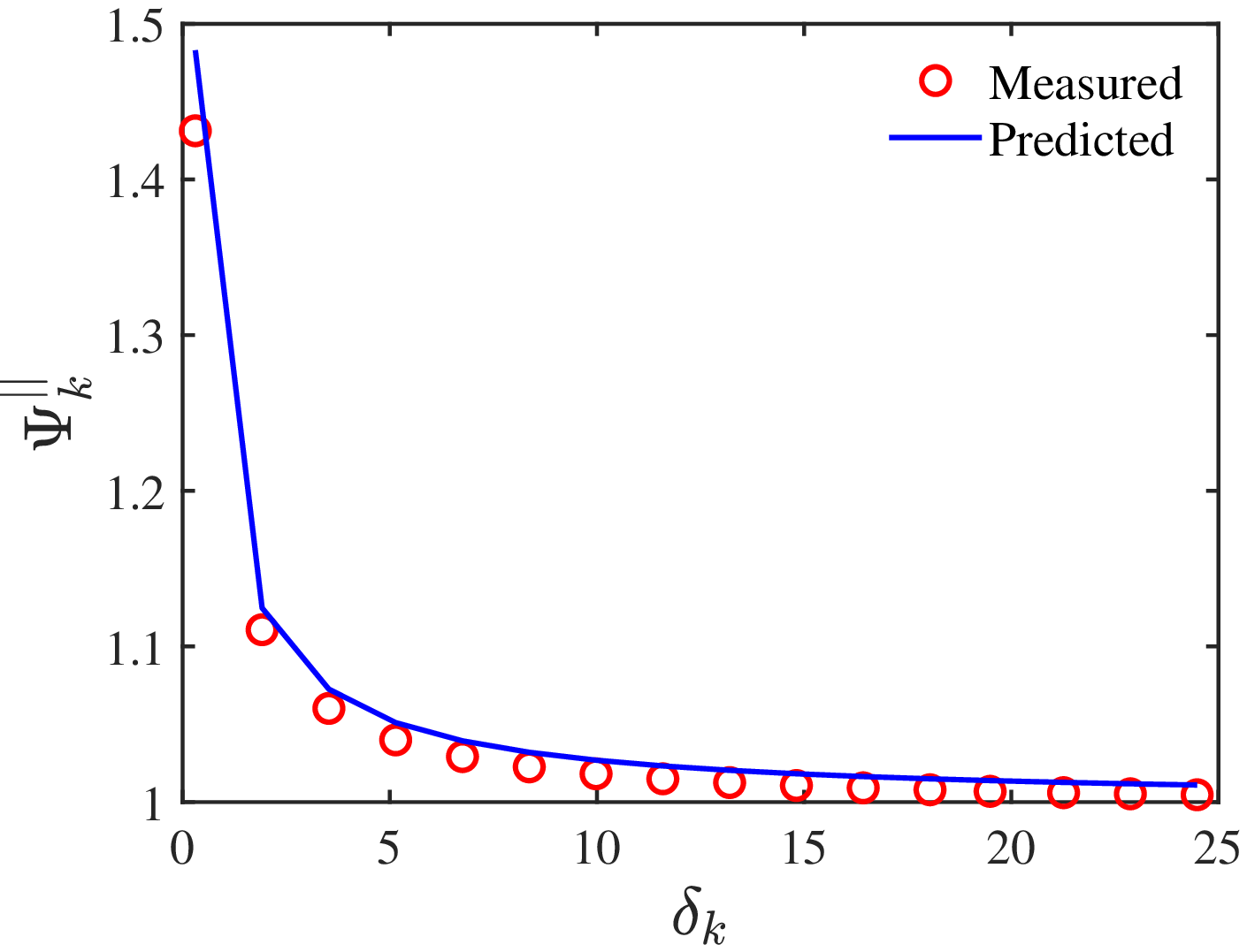}\hspace{0.03\textwidth}
    \includegraphics[scale=0.55]{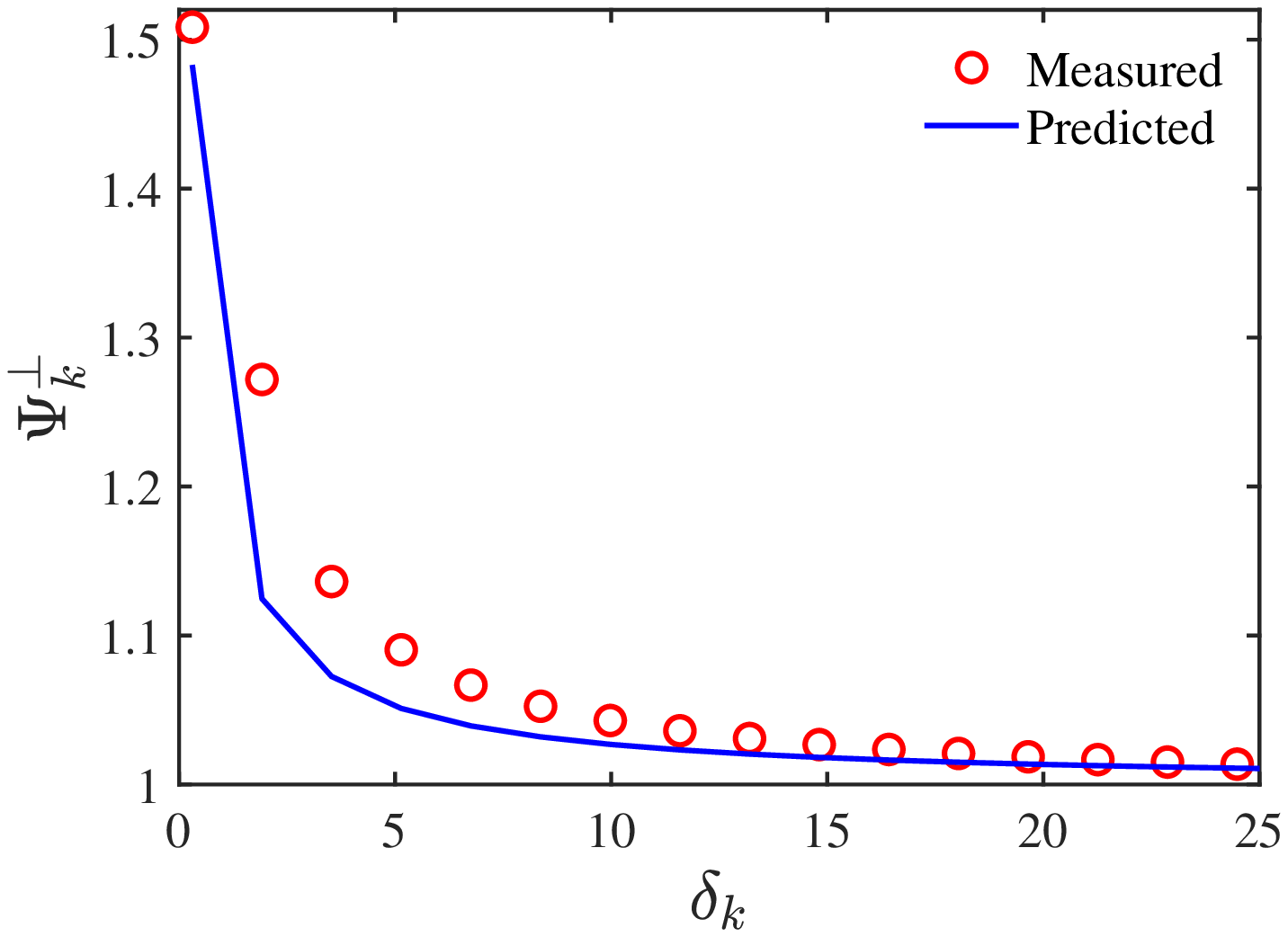}
    \caption{Shown are the predictions of Eq. \ref{eqn:psi_zeng} for the wall adjustment to the drag coefficient of an isotropic computational cell compared to the measured values for parallel (left) and perpendicular (right) forces to the wall.}
    \label{fig:psi_iso}
\end{figure}

%%%%%%%%%%%%%%%%%%
\label{sec:appendix}
A simplified expression for $\Psi^{(i)}_k$ that is only applicable for isotropic grids is introduced here. The new equation denoted by $\Psi^{iso}_k$ is obtained based on the work of \cite{zeng2009}. In their work, an expression using fully resolved direct numerical simulation, was empirically derived for the wall adjustment to the drag coefficient of a solid sphere in parallel motion to a no-slip wall. Our results show that their wall adjustment expression matches our measured values for the wall adjustment to the drag coefficient of isotropic computational cells. Accordingly, the new equation for $\Psi^{(i)}_k$ that is only applicable for isotropic computational cells is introduced based on their empirical expression as 

\begin{equation}
    \Psi^{iso}_k = \left(1.028 - \frac{0.07}{1+4\delta^2_k} - \frac{8}{15}\log\left(\frac{270\delta_k}{135+256\delta_k}\right)\right), 
    \label{eqn:psi_zeng}
\end{equation}

\noindent where 

\begin{equation}
    \delta_k=x^{(2)}_k/(0.5d_c)-0.5,
    \label{eq:delta_c}
\end{equation}

\noindent and $x^{(2)}_k$ is the wall-normal distance of the center of the computational cell $k$, normalized by its equivalent radius of $0.5d_c$. The choice of these two parameters ($x^{(2)}_k$ and $0.5d_c$) are slightly changed compared to the original formulation of \cite{zeng2009} in order to produce better predictions. It is also imperative to mention that Eq. \ref{eqn:psi_zeng} covers a wide range of wall distances and approaches unity when the computational cell is sufficiently away from the wall. Figure \ref{fig:psi_iso} shows the predictions of this equation for both parallel and normal directions compared to the measured values. It should be emphasized that unlike the predictive capability of the equation above for the uniform grid resolutions, it deviates significantly for anisotropic grids with high aspect ratios.

%%%%%%%%%%%%%% Appendix B %%%%%%%%%%%%%%%
\section*{Appendix B. Stokeslet solutions}
%%%%%%%%%%%%%%%%%%%%%%%%%%%%%%%%%%%%%%%%%
In this Appendix, the wall-bounded and unbounded Stokeslet solutions used in the derivation of $\Phi^{(i)}_{kj}$ in section \ref{sec:scheme}, are explained in detail. The unbounded Stokeslet solution that is the flow generated by a point force in an unbounded quiescent fluid with dynamic viscosity of $\mu$ is expressed as \citep{blake1971}

\begin{equation}
u^{(i)}_{stkl,un} = \frac{F^{(j)}}{8\pi\mu}\left( \frac{\delta_{ij}}{|r_{kj}|} + \frac{r^{(i)}_{kj}r^{(j)}_{kj}}{|r_{kj}|^3}\right)
\label{eq:stokeslet_unbounded}
\end{equation}

\noindent where 

\begin{equation}
    r^{(i)}_{kj} = (x^{(i)}_j-x^{(i)}_k) \text{,} \quad |r_{kj}| = \sqrt{ \sum_{i=1}^{3}(r^{(i)}_{kj})^2}
\end{equation}

\noindent and $u^{(i)}$ is the $i$ component of velocity created at the location of $(x^{(1)}_j,x^{(2)}_j,x^{(3)}_j)$ by the point force, $F^{(j)}$, exerted in $j$ direction and located at $(x^{(1)}_k,x^{(2)}_k,x^{(3)}_k)$. $\delta_{ij}$ is the Kronecker delta which is unity for $i{=}j$ and zero otherwise. Similar to this, the wall-bounded Stokeslet solution for a point force that is applied near a no-slip wall is expressed as \citep{blake1971} 

\begin{equation}
    \begin{aligned}
    u^{(i)}_{stkl,b} = & \frac{F^{(j)}}{8\pi\mu} \left[ \left(\frac{\delta_{ij}}{|r_{kj}|} + \frac{r^{(i)}_{kj}r^{(j)}_{kj}}{|r_{kj}|^3} \right) - \left( \frac{\delta_{ij}}{|R_{kj}|} + \frac{R^{(i)}_{kj}R^{(j)}_{kj}}{|R_{kj}|^3} \right) + \right.
    \\ & \left. 2x^{(2)}_k\left( \delta_{jm}\delta_{ml} - \delta_{j3}\delta_{3l}\right)\frac{\partial}{\partial R^{(l)}_{kj}}\left( \frac{x^{(2)}_kR^{(i)}_{kj}}{|R_{kj}|^3}- \left( \frac{\delta_{i3}}{|R_{kj}|} + \frac{R^{(i)}_{kj}R^{(2)}_{kj}}{|R_{kj}|^3}\right)\right) \right]
    \end{aligned}
    \label{eq:stokeslet_bounded}
\end{equation}

\noindent where 

\begin{equation}
    R^{(i)}_{kj} = 
    \begin{cases}
    r^{(i)}_{kj}, \quad i=1,3\\
    r^{(2)}_{kj}+2x^{(2)}_k, \quad i=2
    \end{cases}
    \text{,} \quad |R_{kj}| = \sqrt{ \sum_{i=1}^{3}(R^{(i)}_{kj})^2}
\end{equation}

\noindent and $x^{(2)}_k$ is the wall distance at which the force is applied. The rest of parameters are similar to those of the unbounded Stokeslet solution.

\bibliographystyle{elsarticle-harv}\biboptions{authoryear}
\bibliography{manuscript}

\end{document}